\documentclass[english, letterpaper, 11pt]{article}

\usepackage{fourier}
\usepackage[T1]{fontenc}

\usepackage[margin=1in]{geometry}
\usepackage{amssymb}
\usepackage{amsmath} % For theorems
\usepackage{mathrsfs}
\usepackage{bbm} % For double-stroke unity
\usepackage{scrextend} % For \addmargin{}
\usepackage{natbib} 
\usepackage{graphicx} % Allows inclusion of figures
\usepackage{color}
\usepackage{booktabs} % Allows the use of \toprule, \midrule and \bottomrule in tables
\usepackage[labelsep=period]{caption} % Allows `\\' in captions
\usepackage[normalem]{ulem} % Allows crossing out text
\usepackage[hang,flushmargin]{footmisc}
\usepackage{rotating} % Allows landscape-oriented tables
\usepackage{scalefnt} % Allows scaling fontsize
\usepackage{setspace} % Allows doublespacing
\usepackage{cite}
\usepackage{threeparttable}
\usepackage{caption}
\usepackage{multirow}
\usepackage[hidelinks]{hyperref}
\usepackage{soul}
\usepackage[normalem]{ulem}
\usepackage{authblk} % Author affiliation package

\numberwithin{equation}{section}
\allowdisplaybreaks

\newsavebox{\overlongequation}

\begin{document}
	
\title{\huge Off-Balance Sheet Activities and Scope Economies in U.S.~Banking\thanks{Emir Malikov would like to acknowledge financial support from the Troesh Center for Entrepreneurship \& Innovation at UNLV. \newline\newline  \textit{Email}: jzz0080@auburn.edu (Zhang) and emir.malikov@unlv.edu (Malikov).}}
\author[1]{\sc \vspace{0.2cm} Jingfang Zhang}
\author[2]{\sc  Emir Malikov}
\affil[1]{\small Auburn University}
\affil[2]{\small University of Nevada, Las Vegas}

\date{\small November 10, 2021}
\maketitle

\vspace{\baselineskip}

\begin{abstract}
	\noindent Propelled by the recent financial product innovations involving derivatives, securitization and mortgages, commercial banks are becoming more complex, branching out into many ``nontraditional'' banking operations beyond issuance of loans. This broadening of operational scope in a pursuit of revenue diversification may be beneficial if banks exhibit scope economies. The existing (two-decade-old) empirical evidence lends no support for such product-scope-driven cost economies in banking, but it is greatly outdated and, surprisingly, there has been little (if any) research on this subject despite the drastic transformations that the U.S.~banking industry has undergone over the past two decades in the wake of technological advancements and regulatory changes. Commercial banks have significantly shifted towards nontraditional operations, making the portfolio of products offered by present-day banks very different from that two decades ago. In this paper, we provide new and more robust evidence about scope economies in U.S.~commercial banking. We improve upon the prior literature not only by analyzing the most recent  data and accounting for bank's nontraditional off-balance sheet operations, but also in multiple methodological ways. To test for scope economies, we estimate a flexible time-varying-coefficient panel-data quantile regression model which accommodates three-way heterogeneity across banks. Our results provide strong evidence in support of significantly positive scope economies across banks of virtually all sizes. Contrary to earlier studies, we find no empirical corroboration for scope diseconomies. 
	
	\vspace{\baselineskip}
	
	
	
\end{abstract}

\onehalfspacing
\thispagestyle{empty} \addtocounter{page}{0}
\setlength{\abovedisplayskip}{5pt}
\setlength{\belowdisplayskip}{5pt}
\clearpage

% ------------------------------------------------------------------------------------------
% ------------------------------------------------------------------------------------------	

\section{Introduction}

Just like in other industries, executive managers in banking must choose the optimal scope of operations. Despite the long-lasting implications of this strategic choice for firm performance, the dichotomy between operational ``focus'' and breadth remains unsettled from the corporate strategy perspective. The common arguments for limited-scope operations \`a la \citet{skinner1974decline,skinner1974focused} usually feature cost and quality benefits associated with more specialized expertise and tacit knowledge, lessened complexity, diminished technological uncertainty, etc. On the other hand, there may be a strong incentive to diversify revenue streams by broadening the firm's product mix in order to capitalize on potential scope-driven cost savings and thereby increase firm value \citep[see][]{panzar1981economies,rumelt1982diversification,villalonga2004}. When it comes to commercial banking, leveraging operational scope and breadth thereof continues to play a vital role in operations management. 

The scope of bank operations has also been a subject of intense policy debate, thereby expanding  practical importance of understanding the relation between operational scope and bank performance beyond industry managers and stakeholders. Namely, the financial crisis of 2007--2008 and the ensuing Great Recession turned attention of policy-makers and academics alike  onto large ``too-big-to-fail'' (TBTF) commercial banks and the serious systemic risks that they pose. The emergence of behemoth banks due to deregulation as well as technological innovations (including those in information technologies) has given rise to concerns about the costs that such ``systemically important financial institutions'' impose on the economy and fueled policy debates about whether banks should be subject to size limitations, even including the talks of break-up. These policy discussions have led to  the enactment of new financial regulations such as the Dodd–Frank Wall Street Reform and the Consumer Protection Act of 2010 that seek to eliminate the TBTF doctrine by setting restrictions on the scale and scope of bank operations. However, the potential cost savings associated with operating at a large scale with a more diversified scope of revenue-generating activities, which are to be forgone owing to the new regulations, have been by and large neglected in these policy discussions. 

Large banks may derive such cost efficiency benefits from their ability to offer financial services at lower average cost due to (\textit{i}) ``scale economies'' driven by the increasing returns to scale as well as (\textit{ii}) their unique position to innovate and expand the scope of offered financial products and thereby economize costs (``scope economies'') via input complementarities and positive spillovers \citep[see][]{markideswilliamson1994,milgromroberts1995} as well as, in the case of commercial banking, risk diversification across different products \citep[e.g.,][]{rossietal2009}. 
In theory, these cost savings are passed onto customers in the form of lower net interest margins. This raises an important policy and research question about significance of the trade-off between lower systemic risk pursued by the newly enacted regulations and the cost savings that banks may be forced to forgo as a result. Both have non-negligible implications for consumer welfare. It is therefore imperative to investigate the prevalence of scale and scope economies in banking in order to not only shed light on potential unintended consequences of the financial reforms already put in place but also to inform future policies and regulations. This information also can help banks in formulating optimal product-scope operational strategies.

While studies of scale economies in commercial banking are many, the attempts to measure \textit{scope economies} are however scant and outdated. The latter is especially lacking given the introduction of many ``nontraditional'' financial product innovations involving derivatives, securitization and mortgages by the large banks in the past two decades that have allowed them to expand the scope of their revenue-earning operations. The objective of this paper is to fill in this gap.

Early studies of scale economies in banking date as far back as \citet{berger1987competitive}, \citet{mester1987multiproduct,mester1992traditional} and \citet{hughes1993quality,hughes1998bank} to name a few, and with the passage of new financial reforms, this body of research has only been growing. No matter the methods employed, most recent studies find  empirical evidence in support of the statistically significant increasing returns to scale in the U.S.~banking sector. Some find significant scale economies mostly for large commercial banks \citep[e.g.,][]{wheelock2012large,hughes2013said,restrepo2015nonparametric}; others find economies of scale for medium and small banks as well \citep[e.g.,][]{malikov2015estimation,restrepo2015obelix, wheelock2018evolution}. 

With the sole exception,\footnote{To our knowledge, \citet{yuan2008financial} who explicitly recognize the role of nontraditional banking activities (namely, insurance) is the only attempt at measuring scope economies in the U.S.~banking post 2000. Their analysis looks at a single nontraditional operation and  stops at 2005, which obviously excludes the most relevant period after the structural-change-inducing financial crisis.} there however have been virtually no attempt to investigate product scope economies in banking over the past two decades despite the drastic transformations that this sector has undergone during that time.
This perhaps can be attributed to the lack of empirical evidence in support of statistically and/or economically significant scope economies among U.S.~commercial banks documented in the 1980s and 1990s; e.g., see  \citet{berger1987competitive}, \citet{mester1987multiproduct}, \citet{hughes1993quality}, \citet{pulley1992composite}, \citet{ferrier1993economies}, \citet{pulley1993role}, \citet{jagtiani1995scale}, \citet{jagtiani1996scale},  \citet{wheelock2001new}. It makes scope economies in the present-day banking sector be a seriously overlooked issue because the technological advancements along with regulatory changes have restructured the U.S.~banking industry dramatically, especially since the passage of the Gramm–Leach–Bliley Act in 1999, which largely lifted the restrictions prohibiting the consolidation of commercial banks, investment banks, securities firms and insurance companies. U.S.~banks have since experienced a drastic shift from traditional banking activities (viz., issuance of loans) towards the nontraditional activities such as investment banking, venture capital, security brokerage, insurance underwriting and asset securitization \citep{deyoung2013nontraditional}, and the portfolio of products offered by the modern banks is very different from that two decades ago, underscoring the importance of our study. 

While nontraditional banking operations are usually associated with banks' all other non-interest fee-generating activities related to participating in capital markets, the off-balance sheet banking represents one of the major forms of such nontraditional activities. It chiefly consists of contingent claims/contracts that involve obligations to lend or provide funds should  the contingency be realized and, unlike the traditional interest-income-centered transactions, these off-balance sheet activities are not recorded on the bank’s balance sheet \citep{hassan1993off,hassan1994methodological}.   For example, an interest-earning loan is considered an asset on the bank's balance sheet, whereas a promise to make a loan is an off-balance sheet item since it involves only a \textit{potential} funding obligation in the future, albeit, for which the bank earns a fee. Broadly, off-balance sheet items can be  categorized  into four groups including guarantees, commitments, market-related activities, and advisory or management functions \citep[e.g., see][]{perera2014impact}. Such off-balance sheet banking operations are well-documented to substantially influence banks' financial performance including  profitability and risk profiles \citep[e.g.,][]{stiroh2004diversification,laeven2007there,apergis2014long}, and omitting these revenue-earning operations in the analysis of banking technology may lead to erroneous inference and conclusions due to misspecification \citep[see][]{clark2002x,rime2003performance,casu2005analysis,lozano2010impact}. When testing for scope economies, we therefore recognize off-balance sheet operations as another one of the bank's revenue-generating outputs.

In this paper, we contribute to the literature by providing new and more robust evidence about scope economies in U.S.~commercial banking. We improve upon the prior literature not only by analyzing the most recent and relevant data (2009--2018) and accounting for bank's nontraditional non-interest-centered operations, but also in multiple methodological ways as follows. In a pursuit of robust estimates of scope economies and statistical inference thereon, we estimate a flexible, yet parsimonious, time-varying-coefficient panel-data quantile regression model which accommodates (\textit{i}) distributional heterogeneity in the cost structure of banks along the size of their costs, (\textit{ii}) temporal variation in cost complementarities and spillovers due to technological change/innovation, and (\textit{iii}) unobserved bank heterogeneity (e.g., latent management quality) that, if unaccounted, confounds the estimates. Our analysis is structural in that we explicitly estimate a model of bank cost structure which facilitates the measurement of counterfactual costs necessary to test for scope economies.

By employing a quantile approach, we are able to capture distributional heterogeneity in the bank cost structure. Unlike the traditional regression models that focus on the conditional mean only, quantile regression provides a complete description of the relationship between the distribution of bank costs and its determinants. Since banks of varying size/scale are highly heterogeneous in their operations \citep[e.g., see][]{wheelock2012large}, it is reasonable to expect that large- and small-scale banks exhibit different scope-driven potential for cost saving (if any) and, therefore, there remains much untapped benefit of examining scope economies in banking via quantile analysis. Thus, contrary to all prior studies of scope economies in banking which provide evidence solely for \textit{average} costs via conventional conditional-mean regressions, we focus our analysis on conditional \textit{quantiles} of the bank cost distribution, with the bank's operating cost being a good proxy for its size/scale. Not only does this approach enable us to accommodate potential heterogeneity in the prevalence of scope economies among banks of different sizes, but it is also more robust to the error distributions including the presence of outliers in the data. Furthermore, it exhibits a useful equivariance property thereby letting us avoid biases in the scope economies computations that numerous earlier studies suffer from (to be discussed later).

To operationalize our analysis, we employ the recently developed quantile estimator \citep{machadosantos2019} that we extend to allow temporal variation of unknown form in the parameters in order to flexibly capture the impact of technological innovations on bank operations and costs. Our empirical results provide strong evidence in support of statistically significant scope economies across banks virtually of all sizes in the U.S.~banking sector. Among banks between the bottom 10th and top 90th percentiles of the cost distribution, 92\% or more exhibit positive economies of scope. The prevalence of significant scope economies in median banks is 99\%. Even under the alternative model specifications that produce smaller point estimates, the evidence in support of scope economies in U.S.~banking remains strong, with at least 89\% of mid-cost banks found to enjoy product-scope-driven cost savings. We also find no empirical corroboration for scope \textit{dis}economies. Overall, our findings are in stark contrast with earlier studies. 

The rest of the paper unfolds as follows. Section \ref{sec:scope economies} discusses the theoretical framework. Section \ref{sec:econometrics} describes our econometric model. Data are discussed in Section \ref{sec:data}, followed by  Section \ref{sec:results} that reports the empirical results. We then conclude in Section \ref{sec:conclusion}. 

% ------------------------------------------------------------------------------------------
% ------------------------------------------------------------------------------------------

\section{Theory of Multi-Product Costs}\label{sec:scope economies}

In order to test if there is an untapped cost savings potential for commercial banks due to scope economies, we need to formally model their cost structure. Following the convention in the banking literature, we do so using the dual cost approach. Not only is this approach convenient because it facilitates the direct measurement of the bank's costs via the estimated dual cost function necessary for testing for scope economies, but it also does not require the use of input quantities during the estimation (unlike in the primal production approach) which can lead to simultaneity problems since input allocations are the bank's endogenous decision whereas input prices are widely accepted as being exogenously determined owing to competition in the factor market including that for deposits.

A model of bank costs calls for specification of the outputs and inputs of bank production. 
Given the bank's core functions as a financial intermediary, most studies in the literature adopt \citeauthor{sealey1977inputs}'s (1977) ``intermediation approach'' which focuses on the bank's production of intermediation services and the associated costs inclusive of both the interest and operating expenses. In this paradigm, the revenue-generating financial assets such as loans and trading securities are conceptualized as outputs, whereas inputs are typically specified to include labor, physical capital, deposits and other borrowed funds as well as equity capital \citep[for an excellent review, see][]{hughesmesterforth}. Given the recent industry trends and the growing importance of nontraditional income-earning activities that banks engage in, we also include an output measure of non-interest off-balance sheet income. Together with loans and securities, this  makes a total of $M=3$ outputs.

Concretely, we formalize the bank's cost structure via the following multi-product dual variable cost function:
\begin{equation}\label{eq:costfn_def}
\mathcal{C}_t(\boldsymbol{Y},\boldsymbol{W},\boldsymbol{K}) = \min_{\boldsymbol{X}\ge\mathbf{0}}\left\{ \boldsymbol{X}'\boldsymbol{W}\ |\ (\boldsymbol{X},\boldsymbol{K}) \  \text{can produce}\ \boldsymbol{Y}\ \text{at time}\ t\right\},
\end{equation} 
where the arguments of cost function $\mathcal{C}_t(\cdot)$ are the output quantities $\boldsymbol{Y}\in \Re_+^M $, variable input prices $\boldsymbol{W}\in \Re_+^J $ and fixed input quantities $\boldsymbol{K}\in \Re_+^P $; and  $\boldsymbol{X}\in \Re_+^J $ is the vector of variable input quantities. Importantly, the cost function in \eqref{eq:costfn_def} is time-varying thereby accommodating the evolution of the bank cost structure over time in the face of technological advancements and regulatory changes.

The multi-product firm's cost structure is said to exhibit scope economies if its average cost is decreasing in the number of outputs/operations \citep{panzar1981economies}. Commercial banks may achieve such cost savings by spreading fixed costs (e.g., branch costs and data processing costs) over the more diversified output mix (fixed asset amortization) which now, more often than not, includes nontraditional off-balance sheet operations. Scope economies may also arise from positive spillovers via the (re)use of ``public inputs'' such as client credit information and customer relations as well as intangible assets including tacit knowledge and know-hows. Complementarities across different products can play a big role too. For example, some off-balance sheet operations such as loan commitments (which generate income for banks via fees) essentially represent a technological expansion of traditional lending at a little cost added. At the same time, they can help banks expand the scope of their customer relationship with all the cost-saving informational gains that come with it  \citep{berger1995relationship, das1999theory,degryse2000relationship}.  Banks can also reuse the information gathered when issuing loans to reduce the searching or monitoring requirements of the off-balance sheet activities.

To test for the potential for scope-driven cost savings, we use an expansion-path measure of subadditivity of the bank's cost function \`a la \citet{berger1987competitive}, with the rationale being that subadditivity sheds light on scope economies, the presence of which is a {necessary condition} for the former \citep[see][]{baumoletal1982,evansheckman1984}. Specifically, the subadditivity measure relies on comparison of the costs of smaller \textit{multi}-output banks of \textit{differential} degrees of specialization with the cost of a larger, more diversified bank.\footnote{While preserving the equality of total output quantities on both sides, of course.} Intuitively, this approach zeroes in on scope economies from a perspective of relative\textemdash as opposed to absolute\textemdash notion of revenue diversification. Then, for some distribution weights $0\le\varpi_m^{\kappa}\le 1$ such that $\sum_{\kappa}\varpi_m^{\kappa}=1$ for all $m=1,2,3$ and $\kappa\in\{A,B,C\}$, the bank is said to enjoy scope economies at time $t$ if 
\begin{equation}\label{eq:scope}
\sum_{\kappa\in\{A,B,C\}} \mathcal{C}_t\big( \varpi_1^{\kappa} Y_1,\varpi_2^{\kappa} Y_2,\varpi_3^{\kappa} Y_3 \big)  - \mathcal{C}_t\big( Y_1,Y_2,Y_3 \big) >0,
\end{equation}
where we have suppressed all arguments of the cost function besides outputs.

While the above methodology deviates from the conventional definition of scope economies \citep{baumoletal1982} which relies on the comparison of the cost of producing outputs individually with the cost of their joint production, whereby the bank is said to enjoy scope economies if $\mathcal{C}_t( Y_1,0,0) + \mathcal{C}_t(0, Y_2,0)+ \mathcal{C}_t(0,0, Y_3) - \mathcal{C}_t( Y_1,Y_2,Y_3 ) >0$, it is both more realistic and robust. This is so because it does not require computation of the counterfactual cost of producing
each output separately by a fully specialized \textit{single}-output bank, which naturally suffers from ``excessive extrapolation'' \citep{evansheckman1984,hughes1993quality} since the counterfactuals require extrapolation of the estimated multi-output cost function to its boundaries corresponding to the \textit{non-existent} single-output specializations. Also, the conventional measure of scope economies is just a special case of \eqref{eq:scope} with a pair of weights taking zero values for each counterfactual bank. 

To further avoid excessive extrapolation, we restrict the choice of $\{\varpi_m\}$ to the ``admissible region'' defined by the two data-driven constraints, following \citet{evansheckman1984}. First, each counterfactual bank is ensured to not produce less of each output than banks do in the sample. That is, we require that $\varpi_m^{\kappa} Y_m\ge \min \{Y_m\}$ for all $m=1,2,3$ and $\kappa\in\{A,B,C\}$. The second constraint ensures that each counterfactual bank does not specialize in either one of the outputs to a greater extent than banks do in the sample. In other words, ratios of output quantities for each counterfactual bank must fall in the range of such ratios observed in the data, i.e., for any  pair $Y_m$ and $Y_{m'}$:
\begin{align}\label{eq:scope_constraint}
&\min \left\{\frac{Y_m}{Y_{m'}}\right\} \le \frac{\varpi_m^{\kappa} Y_m^{*}+\min \{Y_m\}}{\varpi_{m'}^{\kappa} Y_{m'}^{*}+\min \{Y_{m'}\}} \le 
\max \left\{\frac{Y_m}{Y_{m'}}\right\},
\end{align}
where $Y_m^{*} = Y_m -3\times\min \{Y_m\}$ for all $m=1,2,3$. Thus, we examine the \textit{within-sample} scope economies.

The quantitative measure of cost subadditivity $\mathcal{S}_t$ (in proportions) is obtained by dividing the expression in \eqref{eq:scope} by $\mathcal{C}_t( Y_1,Y_2,Y_3 )$:
\begin{equation}\label{eq:ES}
\mathcal{S}_t = \frac{ \sum_{\kappa\in\{A,B,C\}} \mathcal{C}_t\Big( \varpi_1^{\kappa} Y_1^*+\min \{Y_1\},\varpi_2^{\kappa} Y_2^*+\min \{Y_2\},\varpi_3^{\kappa} Y_3^*+\min \{Y_3\} \Big) - \mathcal{C}_t\big( Y_1,Y_2,Y_3 \big)}
{ \mathcal{C}_t\big( Y_1,Y_2,Y_3 \big)  },
\end{equation}
where the counterfactual costs under the summation operator have been redefined in order to operationalize the first of the two constraints characterizing the admissible region. Positive (negative) values of $\mathcal{S}_t$ provide evidence of scope economies (\textit{dis}economies); while a zero value suggests scope invariance of the bank's cost structure.

Clearly however, the value of $\mathcal{S}_t$ depends on the choice of distribution weights $\{\varpi_m^{\kappa}\}$. To test for scope economies, we adopt a conservative approach to measuring cost subadditivity, whereby $\{\varpi_m^{\kappa}\}$ are chosen such that the corresponding $\mathcal{S}_t$ is the smallest. With this, ``the'' measure of cost subadditivity (for each bank-year) is
\begin{align}\label{eq:min_ES}
\mathcal{S}_t ^* = \min_{\{\varpi_m^{\kappa}\}}\ \mathcal{S}_t \big(\varpi_m^{\kappa};\ m=1,2,3;\kappa\in\{A,B,C\}\big) \ . 
\end{align}

The rationale is as follows. If the \textit{smallest} subadditivity measure is still positive, then one can quite safely infer that scope economies are locally significant over the bank's feasible output space in a given year. Thus, the main hypothesis of interest is as follows.

\textsc{Hypothesis.}\textemdash{\it Consistent with scope economies at time $t$, the cost subadditivity measure $\mathcal{S}_t ^*>0$.}

% ------------------------------------------------------------------------------------------
% ------------------------------------------------------------------------------------------

\section{Empirical Model}
\label{sec:econometrics}

We estimate the bank's dual variable cost function $\mathcal{C}_t(\cdot)$ at different conditional quantiles of costs. 
Let $C_{it}$ be the variable cost of a bank $i=1,\dots,n$ in year $t=1,\dots,T$ and $\boldsymbol{V}_{it}=(\boldsymbol{Y}_{it}',\boldsymbol{W}_{it}',\boldsymbol{K}_{it}')'$ be the vector of (strictly exogenous) cost-function regressors. We use lower case of $C_{it}$ and $\boldsymbol{V}_{it}$ in the following to denote the log transformations of the variables: e.g., $\boldsymbol{v}_{it}=\ln\boldsymbol{V}_{it}$. Letting the bank's variable cost structure be of the translog\footnote{Quadratic log-polynomial.} form and described by a location-scale model \`a la \citet{koenkerbassett1982} extended to accommodate bank fixed effects and time-varying coefficients, we have
\begin{align}\label{eq:model_l}
c_{it}=\big[\beta_{0}+\beta_{0}^*L(t)\big]+
\big[\boldsymbol{\beta}_{1}+\boldsymbol{\beta}_1^*L(t)\big]'\boldsymbol{v}_{it}+ \tfrac{1}{2}\big[\boldsymbol{\beta}_{2} +\boldsymbol{\beta}_2^*L(t)\big]'\text{vec}\left( \boldsymbol{v}_{it}\boldsymbol{v}_{it}'\right)+
\lambda_{i}+u_{it},
\end{align}
with
\begin{align}\label{eq:model_s}
u_{it}= \Big( \big[\gamma_{0}+\gamma_{0}^*S(t)\big]+ \big[\boldsymbol{\gamma}_{1}+\boldsymbol{\gamma}_1^*S(t)\big]'\boldsymbol{v}_{it}+ \tfrac{1}{2}\big[\boldsymbol{\gamma}_{2}+\boldsymbol{\gamma}_2^*S(t)\big]'\text{vec}\left(\boldsymbol{v}_{it}\boldsymbol{v}_{it}'\right)+
\sigma_{i}\Big)\varepsilon_{it},
\end{align}
where $\left(\beta_0,\boldsymbol{\beta}_{1}',\boldsymbol{\beta}_{2}', \beta_0^*,\boldsymbol{\beta}_{1}^*{'},\boldsymbol{\beta}_{2}^*{'}\right)'$ are unknown location-function coefficients; $\left(\gamma_0,\boldsymbol{\gamma}_{1}',\boldsymbol{\gamma}_{2}', \gamma_0^*,\boldsymbol{\gamma}_{1}^*{'},\boldsymbol{\gamma}_{2}^*{'}\right)'$ are unknown scale-function coefficients; and $\lambda_i$ and $\sigma_i$ are the unobserved bank-specific location and scale fixed effects, respectively.  

To allow for technological change in the bank cost structure, we borrow from \citet{baltagi1988general} and introduce two scalar time indices $L(t)$ and $S(t)$. Both time indices are unobservable and can be thought of as the unknown functions of time. Such time indices are advantageous over simple trends (including quadratic) in modeling temporal changes because they provide richer variation in the measurement of technological change and much closer approximation to observed temporal changes than do the simple time trends. Note that index $L(t)$ enters the location function non-neutrally, shifting not only the intercept $\beta_{0}+\beta_{0}^*L(t)$ but also the linear $\boldsymbol{\beta}_{1}+\boldsymbol{\beta}_{1}^*L(t)$ and quadratic slopes $\boldsymbol{\beta}_{2}+\boldsymbol{\beta}_{2}^*L(t)$, thereby allowing for flexible locational shifts in the costs over time. Analogous scale changes over time are allowed by means of $S(t)$. In all, by means of the time indices in both the location and scale functions, we are able to accommodate temporal changes in the \textit{entire} conditional cost distribution.

Essentially, our model in \eqref{eq:model_l}--\eqref{eq:model_s} is a generalization of the popular translog cost-function specification, where all parameters now vary with time, the covariates affect not only the location (centrality) but also the scale (variability) of the conditional cost distribution; and the bank fixed effects are both location- and scale-shifting. The two equations together facilitate a quantile analysis of the bank's cost structure. 
Along the lines of \citet{machadosantos2019} upon whom we build our estimation procedure, we assume that {(\textit{i})} $\varepsilon_{it}$ is $i.i.d.$~across $i$ and $t$ with some cdf $F_{\varepsilon}$; {(\textit{ii})} $\varepsilon_{it}\perp \boldsymbol{v}_{it}$ with the normalizations that $\mathbb{E}\left[\varepsilon_{it}\right]=0$ and $\mathbb{E}\left[|\varepsilon_{it}|\right]=1$; and {(\textit{iii})} $\Pr\left[ \big[\gamma_{0}+\gamma_{0}^*S(t)\big]+ \big[\boldsymbol{\gamma}_{1}+\boldsymbol{\gamma}_1^*S(t)\big]'\boldsymbol{v}_{it}+ \tfrac{1}{2}\big[\boldsymbol{\gamma}_{2}+\boldsymbol{\gamma}_2^*S(t)\big]'\times\right.$ $\left. \text{vec}\left(\boldsymbol{v}_{it}\boldsymbol{v}_{it}'\right)+
\sigma_{i}>0\right]=1$. Then, for any given quantile index $\tau\in(0,1)$, the $\tau$th conditional quantile function of the log-cost $c_{it}$ implied by \eqref{eq:model_l}--\eqref{eq:model_s} is 
\begin{align}\label{eq:model_quant}
\mathcal{Q}_{c}\left[\tau|\boldsymbol{v}_{it}\right]=&\ 
\overbrace{\Big[{\beta}_{0}+{\gamma}_{0}{q}_{\tau}+{\beta}_{0}^*L(t)+{\gamma}_{0}^*S(t)q_{\tau}\Big]}^{t\text{-varying quantile intercept}}+
\overbrace{\Big[\boldsymbol{\beta}_{1}+\boldsymbol{\gamma}_{1}q_{\tau}+\boldsymbol{\beta}_1^*L(t)+\boldsymbol{\gamma}_1^*S(t)q_{\tau}\Big]'}^{t\text{-varying linear quantile slopes}}\boldsymbol{v}_{it}\ + \notag \\ 
&\ \frac{1}{2}\underbrace{\Big[\boldsymbol{\beta}_{2}+{\boldsymbol{\gamma}}_{2}{q}_{\tau}+\boldsymbol{\beta}_2^*L(t)+\boldsymbol{\gamma}_2^*S(t)q_{\tau}\Big]'}_{t\text{-varying quadratic quantile slopes}}\text{vec}\left( \boldsymbol{v}_{it}\boldsymbol{v}_{it}'\right) +\underbrace{\Big[\lambda_{i}+\sigma_iq_{\tau}\Big],}_{\text{individual quantile fixed effect}} 
\end{align}
where $q_{\tau}=F_{\varepsilon}^{-1}(\tau)$ is the (unknown) $\tau$th quantile of $\varepsilon_{it}$. 

The translog cost model in \eqref{eq:model_quant} is quantile-specific because all bracketed ``composite'' coefficients vary not only with time but also with the cost quantile $\tau$. Furthermore, the technological change in the cost frontier is also quantile-specific thereby allowing for heterogeneous temporal shifts across the {entire} cost distribution as opposed to a shift in the mean only. The unobserved bank fixed effect inside the last brackets is also quantile-specific. Thus, quantile model \eqref{eq:model_quant} can be  rewritten compactly as
\begin{align}\label{eq:model_quant_equiv}
\mathcal{Q}_{c}\left[\tau|\boldsymbol{v}_{it}\right]\equiv&\ \alpha_{0}(\tau,t) + \boldsymbol{\alpha_{1}}(\tau,t)'\boldsymbol{v}_{it} +
\frac{1}{2}\boldsymbol{\alpha_{2}}(\tau,t)'\text{vec}\left( \boldsymbol{v}_{it}\boldsymbol{v}_{it}'\right) +
\mu_{i,\tau},
\end{align}
with the ``alpha'' coefficients corresponding to the bracketed expressions in \eqref{eq:model_quant} and $\mu_i\equiv\lambda_{i}+\sigma_iq_{\tau}$.

We opt to begin with the location-scale model to derive the conditional quantile function of interest in \eqref{eq:model_quant} as opposed to postulating a quantile regression \`a la \eqref{eq:model_quant_equiv} \textit{prima facie} because we seek to estimate these quantiles \textit{indirectly}. This is motivated by the presence of unobserved fixed effects in the quantile model. Namely, since there is no known general transformation that can purge unit fixed effects from the quantile model (owing to nonlinearity of the quantile operator), in such a case the routine check-function-based estimators proceed to \textit{directly} estimate a vector of individual effects by means of including a full set of unit dummies. However, as noted by \citet{koenker2004}, the introduction of a large number of unit fixed effects significantly inflates the variability of estimates of the main parameters of interest, i.e., the slope coefficients. Furthermore, the optimization of an $L_1$-norm corresponding to the check-function-based estimators, when there is a large number of binary variables and the associated parameters to be estimated, is well-known to be computationally cumbersome and oftentimes intractable in practice.\footnote{For instance, in our empirical application $n>7,500$.} The traditional solution to this  assumes that unit fixed effects are only location-shifting and regularizes these individual effects by shrinking them to a common value \citep[see][]{koenker2004,lamarche2010}, but these estimators have gained little popularity in applied work largely because of their complexity. While there is an alternative fixed-effect quantile estimator proposed by \citet{canay2011} that requires no regularization and is notably simpler to implement, it continues to assume that the unit fixed effects have a pure location shift effect. Using the notation of \eqref{eq:model_quant_equiv}, this is tantamount to assuming that $\mu_{i,\tau}=\mu_{i}$ for all $\tau$. Furthermore, none of these check-function-based estimators guarantee that the estimates of regression quantiles do not cross, which is a pervasive but oft-ignored problem in applied work. We therefore adopt the approach recently proposed by \citet{machadosantos2019} that allows an easy-to-implement \textit{indirect} estimation of the quantile parameters via moments, where all parameters are estimated based on the moments implied by the location-scale model in \eqref{eq:model_l}--\eqref{eq:model_s}. Besides its relative computational simplicity, this approach is advantageous for its ability to control for unobserved unit heterogeneity that is both location- and scale-shifting: the individual effects are allowed to affect the entire distribution rather than just shifting its location (therefore, $\{\mu_{i,\tau}\}$ are also quantile-specific). Lastly but not least importantly, this moment-based approach can be easily applied to nonlinear-in-parameters models (like ours is) and produces non-crossing quantile regressions.

To operationalize the estimator, we model unobservable $L(t)$ and $S(t)$ via discretization. For each $\kappa=1,\dots,T$, define the dummy variable $D_{\kappa,t}$ that is equal to 1 in the $\kappa$th time period and 0 otherwise. Then, we discretize time indices as $L(t)=\sum_{\kappa=2}^{T}\eta_{\kappa}D_{\kappa,t}$ and $S(t)=\sum_{\kappa=2}^{T}\theta_{\kappa}D_{\kappa,t}$, where $L(1)=\eta_{1}=0$ and $S(1)=\theta_{1}=0$ are normalized for identification. Parameter identification also requires that both $\beta_0^*$ and $\gamma_0^*$ be normalized; we set $\beta_0^*=\gamma_0^*=1$. Under these identifying normalizations, $\beta_0$, $\boldsymbol{\beta}_1$, $\gamma_{0}$ and $\boldsymbol{\gamma}_1$ are naturally interpretable as ``reference'' coefficients in time period $t=1$. Then, a feasible analogue of the $\tau$th conditional cost quantile in \eqref{eq:model_quant} is given by
\begin{align}\label{eq:model_quant_feasible}
Q_{c}\left[\tau|\boldsymbol{v}_{it}\right]=&\ \Big[{\beta}_{0}+{\gamma}_{0}{q}_{\tau}+ \sum_{\kappa}(\eta_{\kappa}+\theta_{\kappa}q_{\tau})D_{\kappa,t}\Big]+
\Big[\boldsymbol{\beta}_{1}+\boldsymbol{\gamma}_{1}q_{\tau}+\sum_{\kappa}(\boldsymbol{\beta}_1^*\eta_{\kappa}+\boldsymbol{\gamma}_1^*\theta_{\kappa}q_{\tau})D_{\kappa,t}\Big]'\boldsymbol{v}_{it}\ + \notag \\ 
&\ \frac{1}{2}\Big[\boldsymbol{\beta}_{2}+{\boldsymbol{\gamma}}_{2}{q}_{\tau}+\sum_{\kappa}(\boldsymbol{\beta}_2^*\eta_{\kappa}+\boldsymbol{\gamma}_2^*\theta_{\kappa}q_{\tau})D_{\kappa,t}\Big]'\text{vec}\left( \boldsymbol{v}_{it}\boldsymbol{v}_{it}'\right) +\Big[\lambda_{i}+\sigma_iq_{\tau}\Big].
\end{align}

Two remarks are in order. First, the discretized parameterization of the unknown $L(t)$ and $S(t)$ is akin to a nonparametric local-constant estimation of these unknown functions of time with the bandwidth parameter being set to 0. Second, though it might appear at first that, when $L(t)$ and $S(t)$ are modeled using a series of time dummies, we obtain the time-varying slope coefficients on $\boldsymbol{v}_{it}$ by merely interacting the latter with time dummies and adding them as additional regressors, this is \textit{not} the case here because time dummies are restricted to have the same parameters $\{\eta_{k}\}$ and $\{\theta_{k}\}$ both when entering additively as well as when interacting with $\boldsymbol{v}_{it}$. Thus, the location and scale functions are not ``fully saturated'' specification but, in fact, are more parsimonious \textit{nonlinear} (in parameters) functions with much fewer unknown parameters. In avoiding a fully saturated specification that is equivalent to sample-splitting into cross-sections, we accommodate time-invariant bank fixed effects.

% ------------------------------------------------------------------------------------------

\subsection{Estimation Procedure}

Although the estimation of \eqref{eq:model_quant} [or \eqref{eq:model_quant_feasible}] can be done in one step via nonlinear method of moments, we adopt a multi-step procedure that is significantly easier to implement. This is possible because the moments implied by model \eqref{eq:model_l}--\eqref{eq:model_s} and its assumptions are sequential in nature. In other words, we can first estimate parameters of the location function and then those of the scale function in two separate steps. After that, based on the estimates of these parameters, the third step is taken to estimate unknown quantiles and, ultimately, recover time-varying quantile coefficients in \eqref{eq:model_quant}. In what follows, we briefly describe this procedure, with more details available in Appendix \ref{sec:appx_estimator}.

\textbf{Step 1.} We first estimate parameters of the location function. For ease of notation, let $\boldsymbol{D}_{t}=[D_{2,t},\dots,D_{T,t}]'$ and $\boldsymbol{\eta}=[\eta_2,\dots,\eta_T]'$. Under the assumption (\textit{ii}), from \eqref{eq:model_l} it follows that the conditional mean function of the log-cost $c_{it}$ is  
\begin{equation}\label{eq:model_step1}
\mathbb{E}\left[c_{it}| \boldsymbol{v}_{it},\boldsymbol{D}_t\right]=\beta_{0}+\boldsymbol{\eta}'\boldsymbol{D}_t+\Big[\boldsymbol{\beta}_{1}+\boldsymbol{\eta}'\boldsymbol{D}_t\cdot\boldsymbol{\beta}_1^*\Big]'\boldsymbol{v}_{it}+ \frac{1}{2}\Big[\boldsymbol{\beta}_{2}+\boldsymbol{\eta}'\boldsymbol{D}_t\cdot\boldsymbol{\beta}_2^*\Big]'\text{vec}\left( \boldsymbol{v}_{it}\boldsymbol{v}_{it}'\right) +\lambda_{i},
\end{equation}
which can be consistently estimated in the within-transformed form via nonlinear least squares after purging additive location fixed effects. Having obtained the nonlinear fixed-effects estimates of the slope coefficients $\big(\widehat{\boldsymbol{\eta}}',\widehat{\boldsymbol{\beta}}_1',\widehat{\boldsymbol{\beta}}_1^{*}{'},\widehat{\boldsymbol{\beta}}_2',\widehat{\boldsymbol{\beta}}_2^{*}{'}\big)'$, we can then recover the location-shifting intercept ${\beta}_{0}$ and fixed effects $\{\lambda_i\}$ under the usual $\sum_{i=1}^n\lambda_{i}=0$ normalization:
\begin{align}
\widehat{\beta}_{0}&=\frac{1}{nT}\sum_{i}\sum_{t}\Bigg(c_{it}-\widehat{\boldsymbol{\eta}}'\boldsymbol{D}_t -\Big[\widehat{\boldsymbol{\beta}}_{1}+\widehat{\boldsymbol{\eta}}'\boldsymbol{D}_t\cdot\widehat{\boldsymbol{\beta}}_1^*\Big]'\boldsymbol{v}_{it}-
\frac{1}{2}\Big[\widehat{\boldsymbol{\beta}}_{2}+\widehat{\boldsymbol{\eta}}'\boldsymbol{D}_t\cdot\widehat{\boldsymbol{\beta}}_2^*\Big]' \text{vec}\left( \boldsymbol{v}_{it}\boldsymbol{v}_{it}'\right) \Bigg), \label{eq:step1_1} \\
\widehat{\lambda}_{i}&=\frac{1}{T}\sum_{t}\Bigg(c_{it}-\widehat{\beta}_0-\widehat{\boldsymbol{\eta}}'\boldsymbol{D}_t -\Big[\widehat{\boldsymbol{\beta}}_{1}+\widehat{\boldsymbol{\eta}}'\boldsymbol{D}_t\cdot\widehat{\boldsymbol{\beta}}_1^*\Big]'\boldsymbol{v}_{it}-
\frac{1}{2}\Big[\widehat{\boldsymbol{\beta}}_{2}+\widehat{\boldsymbol{\eta}}'\boldsymbol{D}_t\cdot\widehat{\boldsymbol{\beta}}_2^*\Big]' \text{vec}\left( \boldsymbol{v}_{it}\boldsymbol{v}_{it}'\right) \Bigg)\ \forall i. \label{eq:step1_2}
\end{align}
Hence, the residual is $
\widehat{u}_{it}=c_{it}-\widehat{\beta}_0-\widehat{\boldsymbol{\eta}}'\boldsymbol{D}_t -\Big[\widehat{\boldsymbol{\beta}}_{1}+\widehat{\boldsymbol{\eta}}'\boldsymbol{D}_t\cdot\widehat{\boldsymbol{\beta}}_1^*\Big]'\boldsymbol{v}_{it}-
\frac{1}{2}\Big[\widehat{\boldsymbol{\beta}}_{2}+\widehat{\boldsymbol{\eta}}'\boldsymbol{D}_t\cdot\widehat{\boldsymbol{\beta}}_2^*\Big]' \text{vec}\left( \boldsymbol{v}_{it}\boldsymbol{v}_{it}'\right)-\widehat{\lambda}_{i}$.
	
\textbf{Step 2.} We then estimate parameters of the scale function. Based on the assumptions (\textit{ii})--(\textit{iii}), we have an auxiliary conditional mean regression:
\begin{equation}\label{eq:model_2}
\mathbb{E}\left[|u_{it}|| \boldsymbol{v}_{it},\boldsymbol{D}_t\right]=\gamma_{0}+\boldsymbol{\theta}'\boldsymbol{D}_t+\Big[\boldsymbol{\gamma}_{1}+\boldsymbol{\theta}'\boldsymbol{D}_t\cdot\boldsymbol{\gamma}_1^*\Big]'\boldsymbol{v}_{it}+ \frac{1}{2}\Big[\boldsymbol{\gamma}_{2}+\boldsymbol{\theta}'\boldsymbol{D}_t\cdot\boldsymbol{\gamma}_2^*\Big]'\text{vec}\left( \boldsymbol{v}_{it}\boldsymbol{v}_{it}'\right) +\sigma_{i},
\end{equation}
where $\boldsymbol{\theta}=[\theta_2,\dots,\theta_T]'$ and which, just like in the first step, we can estimate via nonlinear least squares after within-transforming scale fixed effects out. This yields the estimates of the scale-function slope coefficients $\big(\widehat{\boldsymbol{\theta}}',\widehat{\boldsymbol{\gamma}}_1',\widehat{\boldsymbol{\gamma}}_1^{*}{'},\widehat{\boldsymbol{\gamma}}_2',\widehat{\boldsymbol{\gamma}}_2^{*}{'}\big)'$. To recover the scale-shifting intercept ${\gamma}_{0}$ and fixed effects $\{\sigma_i\}$, use $\sum_{i=1}^n\sigma_i=0$:
\begin{align}
\widehat{\gamma}_{0}&=\frac{1}{nT}\sum_{i}\sum_{t}\Bigg(|\widehat{u}_{it}|-\widehat{\boldsymbol{\theta}}'\boldsymbol{D}_t -\Big[\widehat{\boldsymbol{\gamma}}_{1}+\widehat{\boldsymbol{\theta}}'\boldsymbol{D}_t\cdot\widehat{\boldsymbol{\gamma}}_1^*\Big]'\boldsymbol{v}_{it}-
\frac{1}{2}\Big[\widehat{\boldsymbol{\gamma}}_{2}+\widehat{\boldsymbol{\theta}}'\boldsymbol{D}_t\cdot\widehat{\boldsymbol{\gamma}}_2^*\Big]' \text{vec}\left( \boldsymbol{v}_{it}\boldsymbol{v}_{it}'\right) \Bigg), \label{eq:step2_1} \\
\widehat{\sigma}_{i}&=\frac{1}{T}\sum_{t}\Bigg(|\widehat{u}_{it}|-\widehat{\gamma}_0-\widehat{\boldsymbol{\theta}}'\boldsymbol{D}_t -\Big[\widehat{\boldsymbol{\gamma}}_{1}+\widehat{\boldsymbol{\theta}}'\boldsymbol{D}_t\cdot\widehat{\boldsymbol{\gamma}}_1^*\Big]'\boldsymbol{v}_{it}-
\frac{1}{2}\Big[\widehat{\boldsymbol{\gamma}}_{2}+\widehat{\boldsymbol{\theta}}'\boldsymbol{D}_t\cdot\widehat{\boldsymbol{\gamma}}_2^*\Big]' \text{vec}\left( \boldsymbol{v}_{it}\boldsymbol{v}_{it}'\right) \Bigg)\ \forall i. \label{eq:step2_2}
\end{align}

\textbf{Step 3.} For any given quantile index $0<\tau<1$ of interest, we next estimate the unconditional quantile of $\varepsilon_{it}$. From \eqref{eq:model_s}, we have the conditional quantile function of $u_{it}$:
\begin{align}
\mathcal{Q}_{u}\left[\tau|\boldsymbol{v}_{it},\boldsymbol{D}_t\right]=
\Big(\gamma_{0}+\boldsymbol{\theta}'\boldsymbol{D}_t+\Big[\boldsymbol{\gamma}_{1}+\boldsymbol{\theta}'\boldsymbol{D}_t\cdot\boldsymbol{\gamma}_1^*\Big]'\boldsymbol{v}_{it}+ \frac{1}{2}\Big[\boldsymbol{\gamma}_{2}+\boldsymbol{\theta}'\boldsymbol{D}_t\cdot\boldsymbol{\gamma}_2^*\Big]'\text{vec}\left( \boldsymbol{v}_{it}\boldsymbol{v}_{it}'\right)+\sigma_i\Big)q_{\tau},
\end{align}
and therefore we can estimate $q_{\tau}$ via the standard quantile regression of $\widehat{u}_{it}$ from Step 1 on $\big(\widehat{\gamma}_{0}+\widehat{\boldsymbol{\theta}}'\boldsymbol{D}_t+\big[\widehat{\boldsymbol{\gamma}}_{1}+\widehat{\boldsymbol{\theta}}'\boldsymbol{D}_t\cdot\widehat{\boldsymbol{\gamma}}_1^*\big]'\boldsymbol{v}_{it}+
\frac{1}{2}\big[\widehat{\boldsymbol{\gamma}}_{2}+\widehat{\boldsymbol{\theta}}'\boldsymbol{D}_t\cdot\widehat{\boldsymbol{\gamma}}_2^*\big]' \text{vec}\left( \boldsymbol{v}_{it}\boldsymbol{v}_{it}'\right)+\widehat{\sigma}_i\big)$ from Step 2, with no intercept. With all unknown parameters now estimated, we can construct the estimator of the feasible analogue of the $\tau$th conditional quantile of the log-cost in \eqref{eq:model_quant_feasible}.

For statistical inference, we use bootstrap. To correct for finite-sample biases, we employ \citeauthor{efron1982}'s (1982) bias-corrected bootstrap percentile confidence intervals. Bootstrap also significantly simplifies testing because, owing to a multi-step nature of our estimator, computation of the asymptotic variance of the parameter estimators is not trivial. Due to the panel structure of data, we use wild residual \textit{block} bootstrap, thereby taking into account the potential dependence in residuals within each bank over time. Details are provided in Appendix \ref{sec:inference}.

% ------------------------------------------------------------------------------------------
% ------------------------------------------------------------------------------------------

\section{Data}
\label{sec:data}

The bank-level data come from the Reports of Condition and Income (the so-called Call Reports) and the Uniform Bank Performance Reports (UBPRs). We obtain annual year-end data for all FDIC-insured commercial banks between 2009 and 2018. As already discussed at length, we focus on the post-financial-crisis period. 

Consistent with the widely accepted \citeauthor{sealey1977inputs}'s (1977) ``intermediation approach'' to formalizing production in banking, we define the bank's cost-function arguments as follows. The two traditional interest-income-centered outputs are $Y_1$ --- total loans, which include real estate loans, agricultural loans, commercial and industrial loans, individual consumer loans and other loans, and $Y_2$ --- total securities, which is the sum of securities held-to-maturity and securities held-for-sale. These output categories are conventional and the same as those considered by, e.g., \citet{koetteretal2012} and \citet{wheelock2020new}. The third output included in our analysis ($Y_3$) measures nontraditional off-balance sheet operations. We use a sum of credit-equivalent measures of the bank's various off-balance sheet operations as a proxy for its involvement in nontraditional activities. Namely, we convert off-balance sheet items into their \textit{credit equivalents} which we determine using credit conversion factors that account for the varying credit risk of different nontraditional banking operations.\footnote{For example, financial standby letter of credit and repo-style transactions have a credit conversion factor of 1, whereas performance standby
letters of credit have a factor of 0.5. The conversion factors come from the Call Reports.}  This facilitates comparability of (traditional) on- and (nontraditional) off-balance sheet activities in the analysis of banking production, which makes it a popular practice in the literature \citep[e.g.,][]{jagtiani1996scale,hughes1998bank,stiroh2000did,clark2002x,berger2003explaining,asaftei2008contribution,hughes2013said,wheelock2020new}. 

More concretely, following  \citet{mccord2014financial} and the FFIEC 041 Reports, we compute $Y_3$ by summing credit-equivalent amounts of all off-balance sheet items. For instance, in 2015--2018, Call Reports define these items as off-balance sheet securitization exposures, financial standby letters of credit, performance standby letters of credit and transaction-related contingent items, commercial and similar letters of credit with an original maturity of one year or less, retained recourse on small business obligations sold with recourse, repo-style transactions, unused commitments excluding unused commitments to asset backed commercial paper conduits, unconditionally cancelable commitments, over-the-counter derivatives, centrally cleared derivatives, and all other off-balance sheet liabilities.

We opt for the credit equivalent of off-balance sheet activities over another popular alternative proxy for banks' nontraditional operations based on net non-interest income \citep[e.g.,][]{deyoung2004,deyoung2013nontraditional,lozano2010impact,davies2014too,malikov2015estimation,wheelock2012large,wheelock2018evolution} because the latter can be negative, which makes it an undesirable measure for one of the bank's outputs \citep[see][]{hughes1998bank}. It is, perhaps, even less suitable a measure for an ``output'' in the structural production analysis because of its fundamental conceptual incongruity with how the bank's other outputs are measured following the convention in the literature: namely, it is based on a ``flow'' (income) data whereas loans $Y_1$ and securities $Y_2$ are the ``stock'' (asset) measures. %Although the bank's non-interest income is heavily influenced by off-balance sheet operations, as \citet{clark2002x} point out, it tends to overestimate the amount of off-balance sheet items because many fees and commissions also come from the on-balance-sheet activities. 
No such issue arises when using credit equivalents of off-balance sheet items. Having said that, we also redo our analysis using this alternative income-based measure of nontraditional activities in one of the robustness checks. In this case, following the literature, the $Y_3$ variable is measured using the total non-interest income (inclusive of the income from fiduciary activities, securities brokerage, investment banking, insurance activities, venture capital and the trading revenue)  minus service charges on deposit accounts.

The three variable inputs are $X_1$ --- physical capital measured by fixed assets, $X_2$ --- labor, measured as the number of full-time equivalent employees, and $X_3$ --- total borrowed funds, inclusive of deposits and federal funds. Their respective prices are $W_1$, $W_2$ and $W_3$, where $W_1$ is measured as the expenditures on fixed assets divided by premises and fixed assets, $W_2$  is computed by dividing salaries and employee benefits by the number of full-time equivalent employees, and $W_3$ is computed as the interest expenses on deposits and fed funds divided by the sum of total deposits and fed funds purchased. Total variable cost $C$ is a sum of expenses on $X_1$, $X_2$ and $X_3$. 

% ---------------------------------------
\begin{table}[t]
\centering
\caption{Data Summary Statistics}\label{tab:datasummary}
\footnotesize
\begin{tabular}{lrrrr}
\toprule
Variables& Mean & 1st Qu. & Median  & 3rd Qu.\\
\midrule
$C $  & 13,602.44  & 2,344.10  & 4,612.10   & 10,066.33  \\
$Y_1$ & 424,480.49 & 55,819.58 & 114,898.89 & 265,677.61 \\
$Y_2$ & 118,621.10 & 13,161.49 & 31,803.01  & 77,360.61  \\
$Y_3$ & 27,304.00  & 659.78   & 2,847.73   & 10,503.06  \\
$W_1$ & 50.79     & 14.83    & 21.27     & 33.85     \\
$W_2$ & 57.86     & 47.00    & 54.34     & 64.90     \\
$W_3$ & 0.82      & 0.40     & 0.66      & 1.09      \\
$K_1$ & 70,383.27  & 9,433.85  & 18,382.47  & 40,467.85  \\
$K_2$ & 0.03      & 0.01     & 0.02      & 0.04      \\
$K_3$ & 1.04      & 0.01     & 0.08      & 0.31     \\
\midrule
\multicolumn{5}{p{8.2cm}}{\scriptsize $C$ -- total variable costs; $Y_1$ -- total loans; $Y_2$ -- total securities; $Y_3$ -- off-balance sheet output measured using credit equivalents; $W_1$ -- price of physical capital; $W_2$ -- price of labor; $W_3$ -- price of financial capital; $K_1$ -- total equity; $K_2$ -- the ratio of nonperforming assets to total assets; $K_3$ --the ratio of loan loss provisions to total assets. Variables $C $, $W_1$,  $W_2$, $Y_1$, $Y_2$, $Y_3$, and $K_1$ are in thousands of real 2005 USD. Variables $W_3$, $K_2$ and $K_3$ are in \%.} \\
\bottomrule[1pt]
\end{tabular}
\end{table}
% ---------------------------------------

We also consider equity capital $K_1$ as an additional input. However, due to the unavailability of the price of equity, we follow \citet{berger2003explaining}  and \citet{feng2010efficiency} in modeling $K_1$ as a quasi-fixed input. The treatment of equity as an input to banking production technology is consistent with \citet{hughes1993quality,hughes1998bank} and \citet{ berger2003explaining} in that banks may use it as a source of loanable funds and thus as a cushion against losses. By including equity $K_1$ in the cost analysis, we are therefore also able to control for the bank's insolvency risk along the lines of  \citeauthor{hughes1998bank}'s (2003) arguments, whereby ``an increase in financial capital reduces the probability of insolvency and provides an incentive for allocating additional resources to manage risk in order to protect the larger equity stake'' (p.314). In effect, conditioning the bank's cost on financial capital also allows controlling for quality of loans since the latter is influenced by risk preferences: as \citet{mester1996} explains, risk-averse bank managers may choose to fund their loans with higher equity-to-deposits ratios (and thus less debt) than a risk-neutral bank would. In our analysis, we also condition the bank's cost on two other proxy measures of output quality reflective of credit risk associated with the likelihood that borrowers default on their loans and accrued interest by failing to make payments as contractually obligated. We include two most commonly used proxies: the ratio of nonperforming assets to total assets $K_2$ \citep[e.g.,][]{hughes2013said,wheelock2018evolution,wheelock2020new} and the ratio of loan loss provision to total {assets} $K_3$ \citep[e.g., see][]{laeven2003,acharyaetal2006,berger2010effects}.\footnote{Although we denote these variables as ``$K$,'' we do \textit{not} conceptualize them as the quasi-fixed input quantities} analogous to $K_1$. Obviously, banks' expectations of credit risk are unobservable but as noted by \citet{berger2010effects}, while the former proxy is an \textit{ex-post} measure of the actual incurred losses from lending, the loan loss provisions can be interpreted as an \textit{ex-ante} measure of the level of expected losses and thus as a proxy for expected quality of assets. Controlling for both when modeling bank costs is imperative because lower-quality assets generally require more resources to manage a higher-level risk exposure thereby raising the costs for banks \citep[see][]{hughes2013said}. Following the literature, we define nonperforming assets as a sum of total loans and lease financing receivables past due 30 days or more and still accruing, total loans and lease financing receivables not accruing, other real estate owned, and charge-offs on past-due loans and leases. The loss provision is measured using the total provision for loan and lease losses.

We exclude observations that have negative/missing values for assets, equity, output quantities and input prices, which are likely the result of erroneous data reporting. This leaves us with an operational sample of 44,704 observations for 7,232 banks. We deflate all nominal variables to the 2005 U.S.~dollars using the consumer price index. Table \ref{tab:datasummary} provides summary statistics for our main variables.

% --------------------------------------------------
\begin{figure}[t]
	\centering
	\includegraphics[scale=0.46]{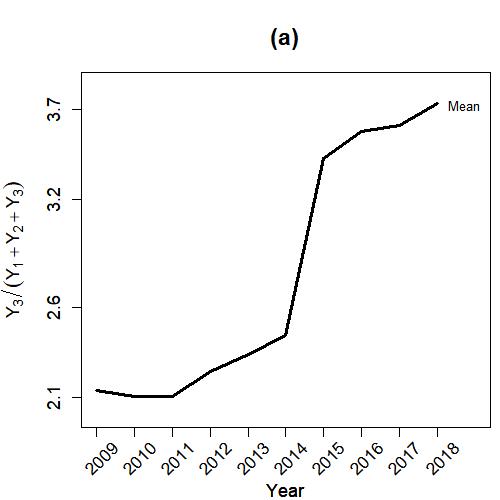}\includegraphics[scale=0.46]{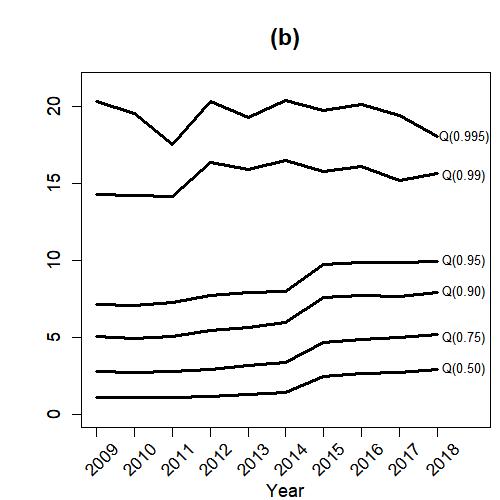}
	\caption{Output Share of Off-Balance sheet Activities over Time, in \%: (a) average, (b) select quantiles}	\label{fig:share} 
\end{figure}
% --------------------------------------------------

Given our emphasis on accounting for banks' off-balance sheet operations, of particular interest is the nontraditional output. Although descriptive statistics in Table \ref{tab:datasummary} expectedly indicate that the volume of $Y_3$ is significantly smaller than that of the two other traditional outputs, banks' involvement in these off-balance sheet activities has, in fact, been steadily expanding in recent years. To show this, we plot the average share of off-balance sheet activities in bank's total output $Y_3/(Y_1+Y_2+Y_3)$ in Figure \ref{fig:share}(a), from where it is evident that the average share of nontraditional outputs among U.S.~commercial banks has been steadily increasing since 2011, almost doubling from about {2}\% in 2009 to {3.7}\% in 2018. This is consistent with the narrative that commercial banks in the U.S.~are increasingly shifting towards off-balance sheet banking. 
	
Obviously, the level of involvement in such nontraditional activities varies considerably across banks, and the rather modest \textit{average} share of the off-balance sheet activities plotted in Figure \ref{fig:share}(a) does not provide a complete picture of the growing prevalence of nontraditional activities in banks' operations because it conceals the well-documented heterogeneity across individual banks. For instance, some banks in our sample are highly specialized in off-balance sheet activities, which account for about 70\% of their outputs. Therefore, we also examine an evolution of the off-balance sheet share in banks' output portfolio at different quantiles in the data, with the particular focus on the upper tail of the distribution. 

% --------------------------------------------------
\begin{figure}[t]
	\centering
	\includegraphics[scale=0.4]{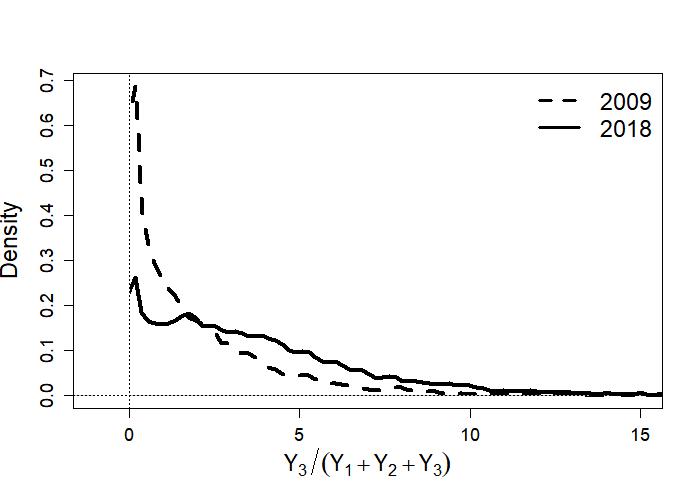}
	\caption{Distribution of the Output Share of Off-Balance Sheet Activities in 2009 vs.~2018}	\label{fig:share_density} 
\end{figure}
% --------------------------------------------------

Figure \ref{fig:share}(b) plots select upper quantiles of $Y_3/(Y_1+Y_2+Y_3)$ over the years, with the lines from bottom to top corresponding to the median, 0.75th, 0.90th, 0.99th and 0.995th quantiles. Two observations are in order here. First, owing to the positive skew in the off-balance sheet share distribution, the differences in banks' involvement in nontraditional banking are stark, with the output share ranging from about 1.8\% at the median to 19.5\% for banks at the top 0.995th quantile. This cross-bank heterogeneity is expected as the choice to engage in nontraditional operations is associated with various idiosyncratic characteristics of banks, including their asset size \citep{rogers1999analysis}. Second, the rising share of off-balance sheet activities is present across just about its entire distribution. The latter is particularly evident in Figure \ref{fig:share_density} that plots this distribution at the beginning (2009) and the end (2018) of our sample period. Altogether, these data document the rising and heterogeneous level of involvement in nontraditional off-balance sheet activities by U.S.~banks in the post-crisis period, not only corroborating the common argument that off-balance sheet activities ought to be accounted in the analysis of banks but also illustrating the importance of adequately accommodating vast heterogeneity across banks in that analysis. We seek to address both these imperatives in our paper.

% ------------------------------------------------------------------------------------------
% ------------------------------------------------------------------------------------------

\section{Empirical Results}
\label{sec:results}

This section reports the results based on our time-varying-coefficient fixed-effects quantile model of bank cost that explicitly accommodates three-way heterogeneity across banks: (\textit{i}) distributional heterogeneity, (\textit{ii}) cross-time heterogeneity and (\textit{iii}) unobserved bank heterogeneity. 

Although our analysis is at different quantiles of the bank's \textit{cost}, the interpretation of distribution heterogeneity can be generalized and extended to bank \textit{size} because the bank's operation cost is a good proxy for its size/scale.
To sufficiently capture distributional heterogeneity across banks, we estimate our model for the $0.10$th, $0.25$th, $0.50$th, $0.75$th and $0.90$th quantiles. The middle three quantiles shed light on the cost structure of mid-size banks in the interquartile range of the conditional log-cost distribution, whereas the more extreme $0.10$th and $0.90$th quantiles provide evidence for the smaller and larger banks, respectively. 

For inference, we use the 95\% bias-corrected bootstrap percentile confidence intervals: one- or two-sided, as appropriate. In what follows, we discuss our main empirical results pertaining to scope economies. We then supplement that discussion by also considering two other sources of potential cost savings in banking, namely, scale economies and technological progress.

% ------------------------------------------------------------------------------------------

\subsection{Scope Economies}

As discussed in Section \ref{sec:scope economies}, we investigate the presence of scope economies by using the expansion-path measure of cost
subadditivity. Since we analyze bank cost structure across the entire cost distribution as opposed to its first moment (i.e., conditional mean), our cost subadditivity measure is not only observation- but also cost-quantile-specific. When evaluating the formulae in \eqref{eq:ES}--\eqref{eq:min_ES}, we replace $\mathcal{C}_{t}(\cdot)$ with the exponentiated quantile function of the log-cost $\mathcal{Q}_c(\tau|\cdot)$ since our cost function estimation is for a conditional log-quantile. That is, for a given quantile $\tau$, we compute the cost subadditivity measure as
\begin{equation}\label{eq:ES_tau}\small
\mathcal{S}_t(\tau) = \frac{ \sum_{\kappa} \exp\left[\mathcal{Q}_c\Big(\tau| \varpi_1^{\kappa} Y_1^*+\min \{Y_1\},\varpi_2^{\kappa} Y_2^*+\min \{Y_2\},\varpi_3^{\kappa} Y_3^*+\min \{Y_3\},t \Big)\right] - \exp\left[\mathcal{Q}_c\big(\tau| Y_1,Y_2,Y_3,t \big)\right]}
{\exp\left[\mathcal{Q}_c\big(\tau| Y_1,Y_2,Y_3,t \big)\right]  }.
\end{equation}

It is noteworthy that our use of quantiles offers another advantage over the more traditional conditional-mean models whereby, owing to a ``monotone equivariance property'' of quantiles, our estimates of $\mathcal{S}_t(\tau)$, which are based on the \textit{level} of cost, are immune to transformation biases due to exponentiation of the estimated \textit{log}-cost function. The same however cannot be said about the estimates of scope economies in analogous conditional-mean analyses. Specifically, to evaluate scope economies, most studies typically exponentiate the predicted \textit{logarithm} of bank cost from the estimated translog conditional-mean regressions while ignoring Jensen's inequality. Consequently, their scope economies estimates are likely biased. To see this, let the conventional fixed-coefficient translog cost regression be $c=f(\boldsymbol{v})+\epsilon$ with $\mathbb{E}[\epsilon|\boldsymbol{v}]=0$, and recall that upper/lower-case variables are in levels/logs. It then trivially follows that $\mathbb{E}[C|\boldsymbol{v}]= \exp\{f(\boldsymbol{v})\}\mathbb{E}[\exp\{\epsilon\}|\boldsymbol{v}]$ which generally diverges from $\exp\{f(\boldsymbol{v})\}$  by a multiplicative function of $\boldsymbol{v}$. Since cost counterfactuals in $\mathcal{S}_t(\tau)$ admit different ``$\boldsymbol{v}$'' values as arguments, the cost subadditivity measure above will normally be biased and need not have the same magnitude or even sign as the true quantity unless $\exp\{\epsilon\}$ is mean-independent of $\boldsymbol{v}$ which is unlikely to be true in practice, say, if $\epsilon$ is heteroskedastic. In the case of quantile estimation, we however do \textit{not} face such a problem owing to the equivariance of quantiles to monotone transformations, viz.~$\mathcal{Q}_{C} [ \tau | \boldsymbol{v} ]=\mathcal{Q}_{\exp\{ c\}} [ \tau | \boldsymbol{v} ] =\exp\{ \mathcal{Q}_c [ \tau | \boldsymbol{v} ]\}$ \citep[e.g., see][]{koenker2005}.

% ---------------------------------------
\begin{table}[t]
\centering
\caption{Cost Subadditivity Estimates}\label{tab:scope}
\footnotesize
\makebox[\linewidth]{
\begin{tabular}{lcccc|cc|cc}
\toprule
Cost & \multicolumn{4}{c}{\textit{Point Estimates}} 	& \multicolumn{4}{c}{\textit{Inference Categories}} \\   
Quantiles ($\tau$) & Mean  & 1st Qu. & Median & 3rd Qu. & $\mathbf{=0}$ & $\ne0$  & $\mathbf{>0}$ & $\le0$ \\
\midrule			
$\mathcal{Q}(0.10)$  & 0.138 & 0.078 & 0.125 & 0.181 & \bf 9.76\% & 90.24\% & \bf 92.04\% & 7.96\% \\
			& (0.058, 0.469) & (0.023, 0.288) & (0.048, 0.463) & (0.082, 0.626) &       &       &    \\
$\mathcal{Q}(0.25)$  & 0.175 & 0.107 & 0.163 & 0.225 & \bf 5.48\% & 94.52\% & \bf 95.70\% & 4.30\% \\
			& (0.078, 0.598) & (0.036, 0.361) & (0.067, 0.579) & (0.106, 0.777) &       &       &     \\
$\mathcal{Q}(0.50)$  & 0.264 & 0.175 & 0.258 & 0.335 & \bf 1.40\% & 98.60\% & \bf 98.90\% & 1.10\%  \\
			& (0.120, 0.937) & (0.066, 0.549) & (0.109, 0.873) & (0.155, 1.185) &       &       &    \\
$\mathcal{Q}(0.75)$  & 0.388 & 0.259 & 0.394 & 0.496 & \bf 0.45\% & 99.55\% & \bf 99.50\% & 0.50\%  \\
			& (0.194, 1.205) & (0.103, 0.683) & (0.169, 1.113) & (0.242, 1.582)  &       &       &     \\
$\mathcal{Q}(0.90)$  & 0.459 & 0.313 & 0.476 & 0.575 & \bf 0.30\% & 99.70\% & \bf 99.60\% & 0.40\%  \\
			& (0.261, 1.164) & (0.121, 0.671) & (0.231, 1.036) & (0.356, 1.567) &       &       &      \\		
\midrule
\multicolumn{9}{p{15.5cm}}{\scriptsize The left panel summarizes point estimates of $\mathcal{S}_t^*(\tau)$ with the corresponding two-sided 95\% bias-corrected confidence intervals in parentheses. Each bank-year is classified as exhibiting scope economies [$\mathcal{S}_t^*(\tau)>0$] vs.~non-economies [$\mathcal{S}_t^*(\tau)\le0$] and scope invariance [$\mathcal{S}_t^*(\tau)=0$] vs.~scope non-invariance [$\mathcal{S}_t^*(\tau)\ne0$] using the corresponding one- and two-sided 95\% bias-corrected confidence bounds, respectively. The right panel reports sample shares for each category and for its corresponding negating alternative. Percentage points sum up to a hundred within binary groups only.} \\
\bottomrule[1pt]
\end{tabular}
}
\end{table}
% ---------------------------------------

Now, recall that $\mathcal{S}_t(\tau)$ depends on the choice of $\{\varpi_m^{\kappa}\}$, which we circumvent by choosing weights that yield the smallest cost subadditivity measure for a given cost quantile $\tau$ in the admissible region: $\mathcal{S}_t^*(\tau)$. Namely, for each fixed cost quantile of interest, we perform a grid search over a permissible range of weights in $[0,1]^6$ at the 0.1 increments. We do this for each bank in a given year. Table \ref{tab:scope} summarizes such point estimates of $\mathcal{S}_t^*(\tau)$ for different quantiles of the conditional cost distribution. (We caution readers against confusing quantiles of the conditional cost distribution $\tau$, for which our bank cost function and the cost subadditivity measure are estimated, with the quantiles of  empirical distribution of observation-specific $\mathcal{S}_t^*(\tau)$ estimates corresponding to a given $\tau$.)

The two hypotheses of particular interest here are (\textit{i}) $\mathbb{H}_0: \mathcal{S}_t^*(\tau)\le0\ \text{v.}\ \mathbb{H}_1: \mathcal{S}_t^*(\tau)>0$ and (\textit{ii}) $\mathbb{H}_0: \mathcal{S}_t^*(\tau)=0\ \text{v.}\ \mathbb{H}_1: \mathcal{S}_t^*(\tau)\ne0$. Both tests are essentially the same, except for the one- or two-sided alternatives. Although the $(i,t)$ index on outputs is suppressed in \eqref{eq:ES_tau}, the tests are at the level of observation (bank-year). In case of (\textit{i}), rejection of the null would imply that even the {smallest} subadditivity measure is statistically \textit{positive} and scope economies can thus be inferred to also be locally significant over the bank's output space in a given year. In case of (\textit{ii}), failure to reject the null would suggest that subadditivity measure is statistically indistinguishable from zero, which is consistent with the bank's cost structure exhibiting local scope invariance.

% --------------------------------------------------
\begin{figure}[p]
	\centering
	\includegraphics[scale=0.29]{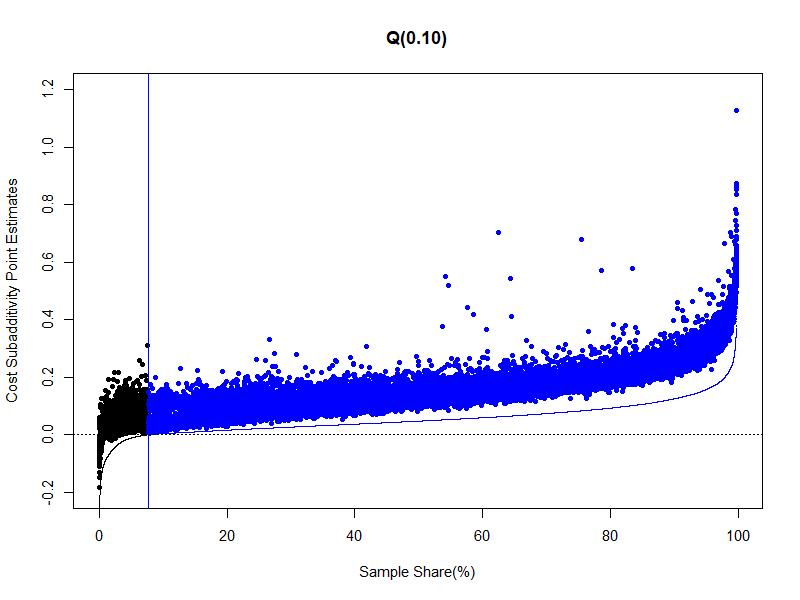}\includegraphics[scale=0.29]{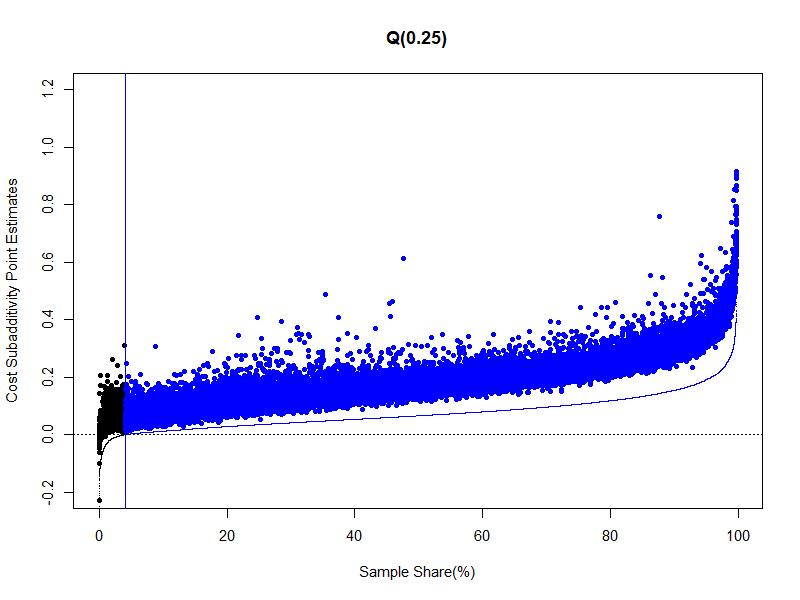}
	\includegraphics[scale=0.29]{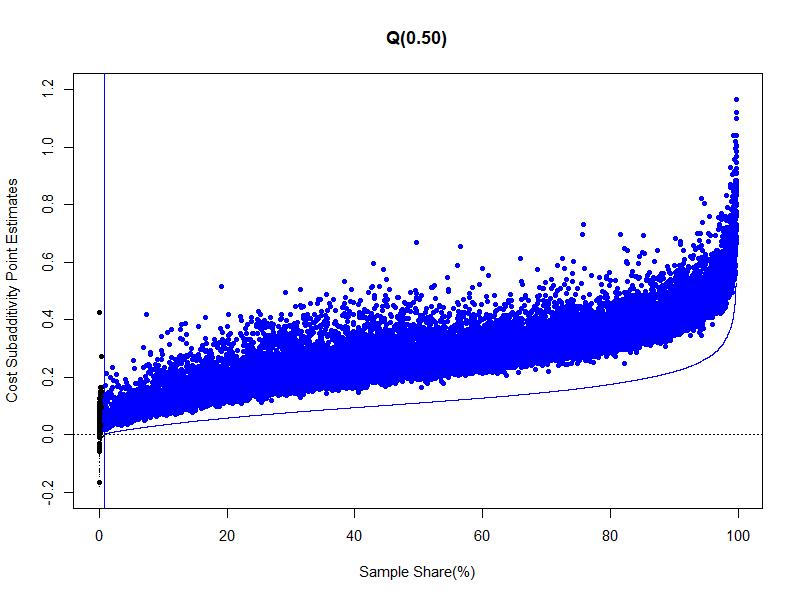}
	\includegraphics[scale=0.29]{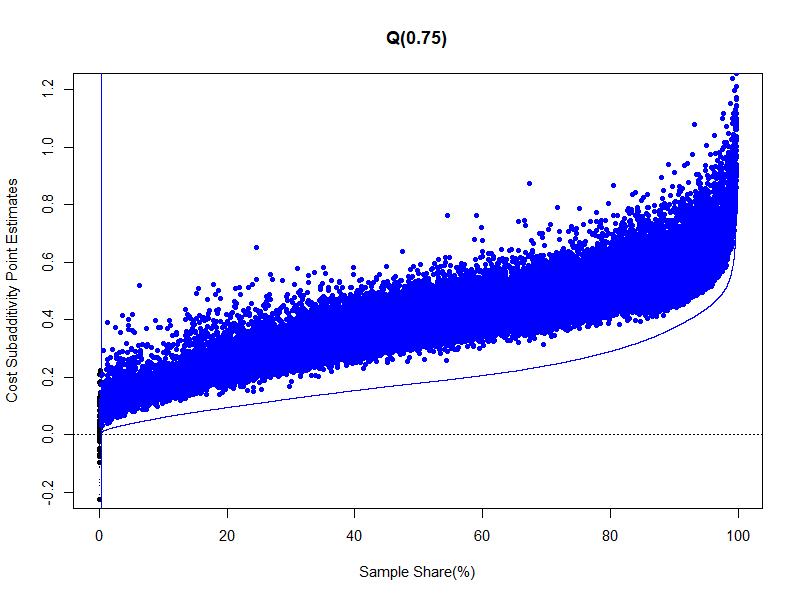}\includegraphics[scale=0.29]{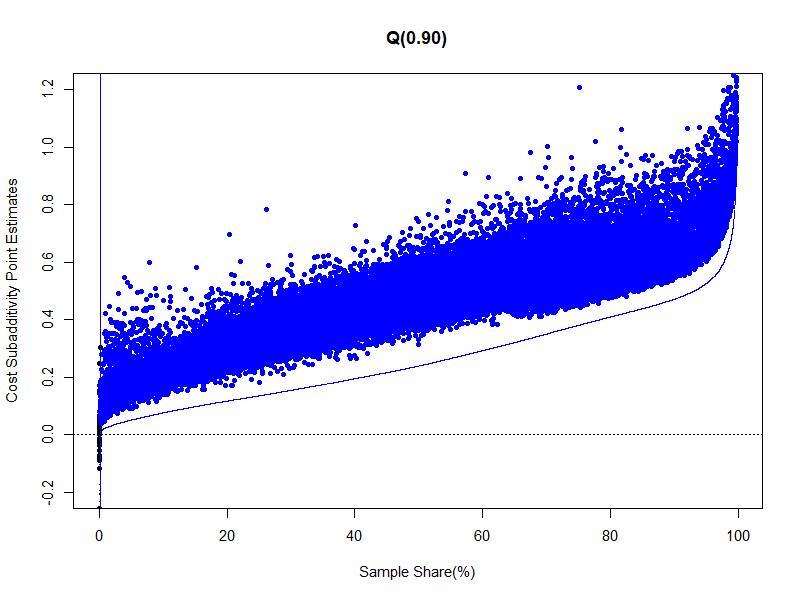}
	\caption{The One-Sided 95\% Lower Bounds (solid lines) of the Cost Subadditivity Point Estimates (scatter points) Across Cost Quantiles} \label{fig:scatterplot}
\end{figure}
% --------------------------------------------------

The right panel of Table \ref{tab:scope} reports the results of these hypothesis tests. Namely, for each cost quantile $\tau$, we classify banks in our data based on the two dichotomous groups of categories: banks that exhibit scope economies [$\mathcal{S}_t^*(\tau)>0$] vs.~scope non-economies [$\mathcal{S}_t^*(\tau)\le0$] and the banks whose cost structure that exhibits scope invariance [$\mathcal{S}_t^*(\tau)=0$] vs.~scope non-invariance [$\mathcal{S}_t^*(\tau)\ne0$]. 
%Note that the two types of categorization are not mutually exclusive by construction.

Our results provide strong evidence in support of statistically significant scope economies across banks virtually of all sizes in the U.S.~banking sector. For banks in the middle interquartile range of the cost\textemdash essentially, size\textemdash distribution, at least 95.7\% exhibit positive economies of scope. For the top half of the distribution (median or higher), the prevalence of significant scope economies is about 99\%. Even at the very bottom of cost distribution ($\tau=0.1$) where the revenue diversification opportunities may not be as abundant or easily accessible, our test results suggest that roughly 92\% of banks enjoy scope-driven cost savings and those, who do not, exhibit scope invariance. Figure \ref{fig:scatterplot} provides a graphic illustration of these results. For each considered cost quantile $\tau$, the figure shows a scatter-plot of the $\mathcal{S}_t^*(\tau)$ estimates for each bank-year observation along with the corresponding one-sided 95\% lower confidence bound. Here, we sort these estimates by their lower confidence bounds (solid line) and color them based on whether they are significantly above 0 or not. From Figure \ref{fig:scatterplot}, it is evident that positive scope economies are ubiquitous and that their presence is only growing with quantiles of the conditional variable-cost distribution of banks (i.e.,  with the bank size).

As a robustness check, we re-estimate our model under alternative empirical specifications of the cost-function variables. Namely, we consider a different proxy for nontraditional operations used in the literature (net non-interest income) as well as assess sensitivity of our findings to credit risk proxies included in the analysis. Table \ref{tab:scope_robust} summarizes estimates of cost subadditivity across these alternatives. Two observations are in order here. First, omitting an \textit{ex-ante} proxy for output quality (loan loss provisions) produces uniformly larger point estimates of cost subadditivity. Consequently, the evidence in favor of significantly positive scope economies is even stronger in the latter case. Nonetheless, we continue to include this important control in our main specification. Second, when using net non-interest income as a proxy measure of nontraditional banking operations, we obtain smaller $\mathcal{S}_t^*(\tau)$ estimates, with the largest differences seen at the bottom tail of costs. But even then, the empirical evidence in support of scope economies across banks is strong. For banks in the middle of the cost distribution (at the conditional median), 89\% exhibit significant economies of scope. The prevalence of product-scope-driven cost savings is even more pervasive ($\ge97.6\%$) for larger banks at higher quantiles. In the case of smaller-scale banks at the bottom 0.10th and 0.25th quantiles, the share of banks that enjoy scope economies\textemdash while smaller\textemdash is nonetheless non-negligible, ranging between 41--65\% and 59--82\%, respectively, depending on the credit risk proxies included in the analysis. The cost structure of the remaining banks is scope-invariant. All in all, our findings of significant scope economies in banking are robust to alternative specifications, and in what follows, we therefore focus on the results from our main specification only.

% ---------------------------------------
\begin{table}[t]
\centering
\caption{Cost Subadditivity Estimates: Robustness to Alternative Variable Specifications}\label{tab:scope_robust}
\footnotesize
\makebox[\linewidth]{
\begin{tabular}{l ccc|ccc|ccc|ccc}
\toprule
& \multicolumn{3}{c}{{(I): Main Specification}} & \multicolumn{3}{c}{{(II)}} & \multicolumn{3}{c}{{(III)}} & \multicolumn{3}{c}{(IV)} \\ 
\midrule
Cost & Median & \multicolumn{2}{c}{\textit{Categories}} & Median & \multicolumn{2}{c}{\textit{Categories}} & Median & \multicolumn{2}{c}{\textit{Categories}} & Median & \multicolumn{2}{c}{\textit{Categories}} \\   
Quantiles ($\tau$) & Est. & $\mathbf{=0}$ & $\mathbf{>0}$ & Est. & $\mathbf{=0}$ & $\mathbf{>0}$ & Est. & $\mathbf{=0}$ & $\mathbf{>0}$ & Est. & $\mathbf{=0}$ & $\mathbf{>0}$ \\
\midrule			
$\mathcal{Q}(0.10)$  & 0.125 & 9.8\% & 92.0\% & 0.182 & 2.1\% & 98.3\% & 0.068 & 58.9\% & 40.8\% & 0.239 & 38.0\% & 65.0\%  \\
$\mathcal{Q}(0.25)$  & 0.163 & 5.5\% & 95.7\% & 0.256 & 0.6\% & 99.4\% & 0.120 & 41.6\% & 59.1\% & 0.282 & 20.3\% & 82.3\%  \\
$\mathcal{Q}(0.50)$  & 0.258 & 1.4\% & 98.9\% & 0.429 & 0.1\% & 99.7\% & 0.226 & 11.4\% & 89.0\% & 0.346 & 3.3\% & 97.5\%  \\
$\mathcal{Q}(0.75)$  & 0.394 & 0.5\% & 99.5\% & 0.543 & 0.1\% & 99.7\% & 0.356 & 2.6\% & 97.6\% & 0.409 & 1.0\% & 99.2\%  \\
$\mathcal{Q}(0.90)$  & 0.476 & 0.3\% & 99.6\% & 0.589 & 0.0\% & 99.7\% & 0.427 & 1.7\% & 98.5\% & 0.447 & 0.8\% & 99.4\%  \\	
\midrule
\scriptsize Nontraditional Output Measure:      &\multicolumn{3}{c}{} &\multicolumn{3}{c}{} &\multicolumn{3}{c}{} &\multicolumn{3}{c}{} \\ 
$\quad$ \scriptsize Credit Equivalents      &\multicolumn{3}{c}{$\checkmark$} &\multicolumn{3}{c}{$\checkmark$} &\multicolumn{3}{c}{} &\multicolumn{3}{c}{} \\
$\quad$ \scriptsize Net Non-Interest Income &\multicolumn{3}{c}{} &\multicolumn{3}{c}{} &\multicolumn{3}{c}{$\checkmark$} &\multicolumn{3}{c}{$\checkmark$}\\ 
\scriptsize Credit Risk Proxies:            &\multicolumn{3}{c}{} &\multicolumn{3}{c}{} &\multicolumn{3}{c}{} &\multicolumn{3}{c}{} \\ 
$\quad$ \scriptsize Nonperforming Assets    &\multicolumn{3}{c}{$\checkmark$} &\multicolumn{3}{c}{$\checkmark$} &\multicolumn{3}{c}{$\checkmark$} &\multicolumn{3}{c}{$\checkmark$} \\
$\quad$ \scriptsize Loan Loss Provision     &\multicolumn{3}{c}{$\checkmark$} &\multicolumn{3}{c}{} &\multicolumn{3}{c}{$\checkmark$} &\multicolumn{3}{c}{} \\ 
\midrule
\multicolumn{13}{p{18.3cm}}{\scriptsize Reported are the median point estimates of $\mathcal{S}_t^*(\tau)$ and shares of the sample for which the estimates are statistically $>0$ (i.e., a bank-year is classified as exhibiting scope economies) and statistically not different from $0$ (i.e., a bank-year is classified as exhibiting scope invariance) at the 95\% level. Because the two hypotheses are tested separately, percentage points need not sum up to a hundred. Specification (I) is our main specification, the complete results for which are reported in Table \ref{tab:scope}.} \\
\bottomrule[1pt]
\end{tabular}
}
\end{table}
% ---------------------------------------

A finding worth emphasizing here is that, having accounted for three-way heterogeneity across banks in a pursuit of robust estimates of bank cost subadditivity, we find \textit{no} empirical evidence in support of scope \textit{dis}economies. This is in stark contrast with earlier studies of scope economies in U.S.~banking \citep[e.g.,][]{berger1987competitive, mester1987multiproduct, hughes1993quality, pulley1992composite, ferrier1993economies, pulley1993role, jagtiani1995scale, jagtiani1996scale, wheelock2001new}. Besides our reliance on the more robust estimation methodology, the qualitative differences between our and prior findings can also be attributed to fundamental changes that the banking sector has undergone in the past two decades characterized by the growing importance of nontraditional banking operations propelled by the financial product innovations.

% --------------------------------------------------
\begin{figure}[t]
	\centering
	\includegraphics[scale=0.29]{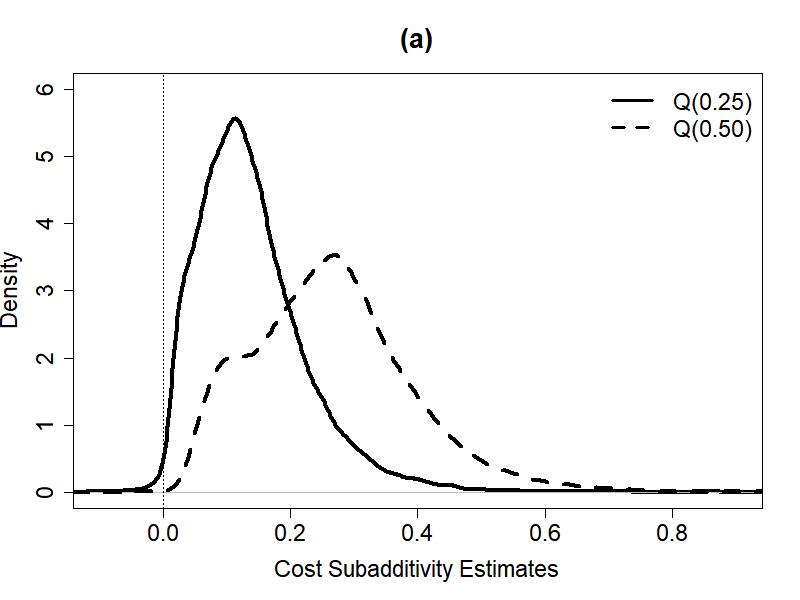}\includegraphics[scale=0.29]{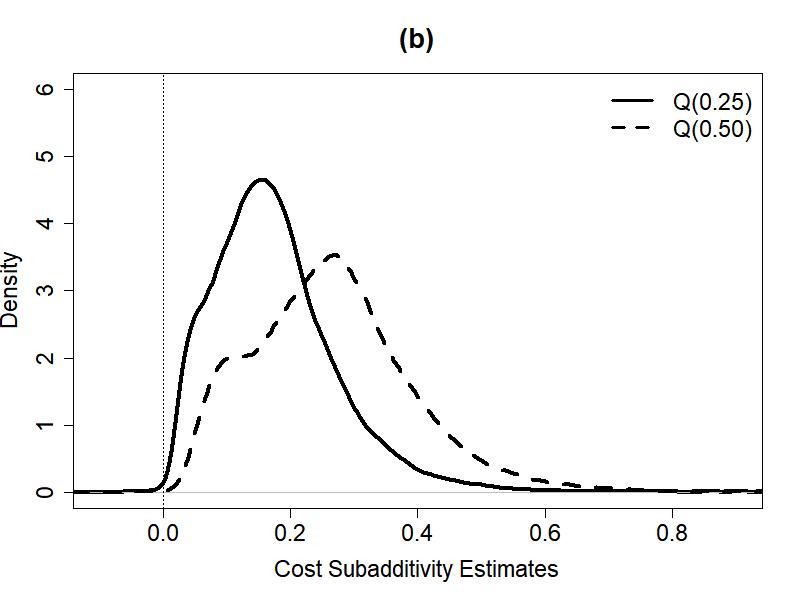}	
	\includegraphics[scale=0.29]{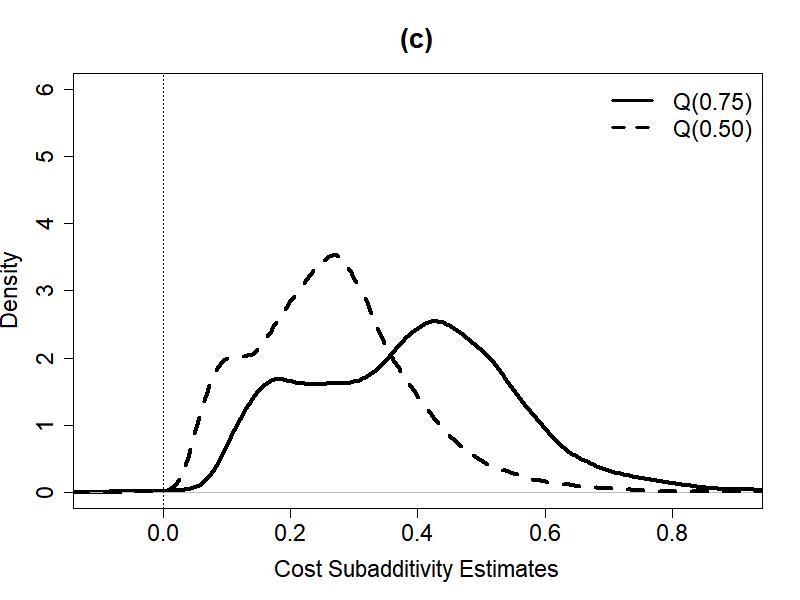}\includegraphics[scale=0.29]{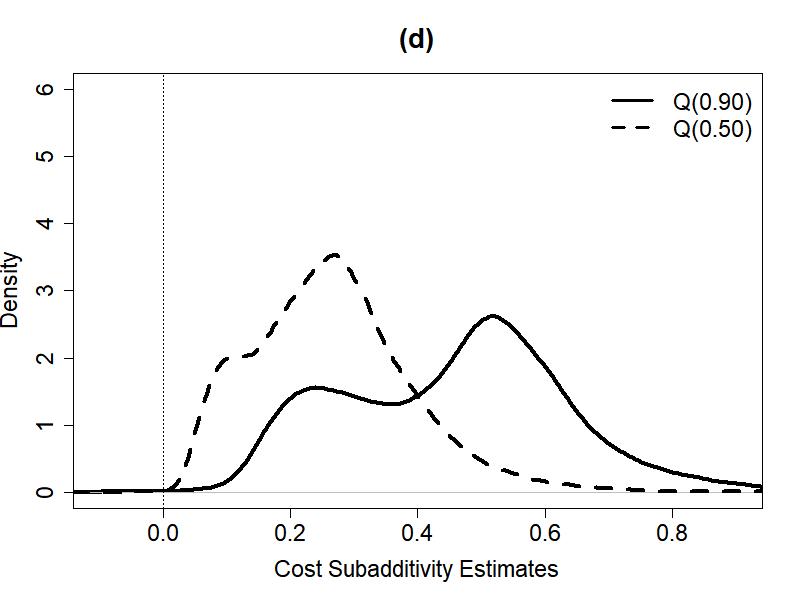}
	\caption{Kernel Densities of Cost Subadditivity Estimates Across Cost Quantiles} \label{fig:densityplot}
\end{figure}
% --------------------------------------------------

Although, the subadditivity measure does not directly quantify the \textit{magnitude} of scope economies in the conventional interpretation of the latter, the value of its point estimates can still provide useful insights into the diversification-driven cost savings. Recall that $\mathcal{S}_t^*(\tau)$ compares the cumulative cost of multiple smaller banks of higher degrees of \textit{relative} output specialization with the cost of a larger, more relatively diversified bank. Essentially, the subadditivity measure sheds light on scope economies from a perspective of relative\textemdash as opposed to absolute\textemdash notion of revenue diversification. Measured is the reduction in bank cost (in proportions) afforded by achieving lower specialization in any one output. From the left panel of Table \ref{tab:scope}, the mean estimates of cost subadditivity ranges from 0.138 to 0.459 depending on the conditional cost quantile. This suggests, on average, the potential for a 14--46\% cost saving if the bank ``rebalances'' its joint production of loans, securities and off-balance sheet outputs. We also find that the magnitude of diversification-driven  economies increases as one moves from the bottom to top of the bank cost distribution, thereby suggesting that larger banks (higher $\tau$) may economize cost better compared to those of smaller size in the lower end of the cost distribution.

For a more holistic look at the empirical evidence of scope economies across different quantiles of the bank cost distribution, we also provide kernel density plots of the $\mathcal{S}_t^*(\tau)$ estimates in Figure \ref{fig:densityplot}. It enables us to compare distributions of the cost subadditivity estimates as opposed to merely focusing on marginal moments. Consistent with our earlier discussion, these plots indicate that large-scale banks lying in the upper quantiles of the cost distribution appear to enjoy bigger diversification-driven cost economies than those in the lower cost quantiles. To support this visual evidence, we formally test for the (first-order) stochastic dominance of scope economies exhibited by banks in the top cost quantiles over those exhibited by those in the bottom quantiles. We utilize a generalized Kolmogorov-Smirnov test proposed by \citet{linton2005consistent} which permits testing dominance over multiple variables (in our case, more than two cost quantiles) and allows these variables to be estimated latent quantities as opposed to observables from the data and to also share dependence (in our case, the dependence is due to common parameter estimates used to construct quantile coefficients). Specifically, let $F_{\tau}(\mathcal{S})$ represent the cumulative distribution functions of the $\mathcal{S}_t^*(\tau)$ estimates for a given cost quantile $\tau$. We then form the null hypotheses that diversification-driven scope economies exhibited by banks in the lower quantiles of the cost distribution are stochastically dominated by those in the upper quantiles of the cost distribution.
More formally, for any cost quantile of interest $\overline{\tau}\in\mathbb{T}$ with $\mathbb{T}=\{0.10,0.25,0.50,0.75,0.90\}$, we are interested in 
\begin{align*}
\mathbb{H}_0: \min_{\tau\ne\overline{\tau}\in\mathbb{T}}\sup_{\mathcal{S}\in\mathbb{S}} \left[ F_{\tau}(\mathcal{S})-F_{\overline{\tau}}(\mathcal{S}) \right]\le 0\ \  \text{v.}\  \  \mathbb{H}_1: \min_{\tau\ne\overline{\tau}\in\mathbb{T}}\sup_{\mathcal{S}\in\mathbb{S}} \left[ F_{\tau}(\mathcal{S})-F_{\overline{\tau}}(\mathcal{S}) \right]>0.
\end{align*}
We use the sub-sampling procedure suggested by \citet{linton2005consistent} to perform the test.\footnote{We employ $199$ equidistant sub-sample sizes $B_n=\{b_1,\dots,b_{199}\}$, where $b_1=[\log\log N]$, $b_{199}=[N/\log\log N]$ with $N=nT$ being the sample size. For each sub-sample size, we get a $p$-value. The reported is the mean of these 199 $p$-values.} 

% ---------------------------------------
\begin{table}[t]
\centering
\caption{Stochastic Dominance of Scope Subadditivity Across Cost Quantiles} \label{tab:pvalue}
\footnotesize
\makebox[\linewidth]{
\begin{tabular}{l cccc}
		\toprule
		& $\{ \mathcal{Q}({0.75}),\dots,\mathcal{Q}({0.10})\}$    & $\{ \mathcal{Q}({0.50}),\mathcal{Q}({0.25}),\mathcal{Q}({0.10})\}$   & $\{ \mathcal{Q}({0.25}),\mathcal{Q}({0.10})\}$ & $\mathcal{Q}({0.10})$ \\
		\midrule
		$\mathcal{Q}({90})$ &  0.578&    0.578&   0.739 & 0.970 \\
		$\mathcal{Q}({75})$ &  & 0.894 & 0.970 & 0.970 \\
		$\mathcal{Q}({50})$ &  &       & 0.970 & 0.970 \\
		$\mathcal{Q}({25})$ &    &       &   & 0.784 \\
\midrule
\multicolumn{5}{p{10cm}}{\scriptsize Reported are the $p$-values.} \\
\bottomrule[1pt]
\end{tabular}
}
\end{table}
% ---------------------------------------

Table \ref{tab:pvalue}  reports $p$-values for the tests of  dominance of $\mathcal{S}_t^*(\tau)$ from the ``row'' quantile over a multi-quantile set of $\mathcal{S}_t^*(\tau)$ from the ``column'' quantiles. All $p$-values are safely greater than the conventional 0.05 level, and we fail to reject the nulls. Combined with the visual evidence from Figure \ref{fig:densityplot}, we can therefore infer that bigger banks in the higher quantiles of the cost distribution exhibit larger scope economies than do smaller banks from the lower cost quantiles for the entire set of observable output mixes. Relatedly, of interest is the relation between the scope economies magnitude and the degree of bank's specialization in nontraditional products. To examine this, for each cost quantile $\tau$ that we consider in our analysis, we run a least-absolute-deviation regression of the $\mathcal{S}_t^*(\tau)$ estimates on the share of off-balance sheet activities in bank's total output $Y_3/(Y_1+Y_2+Y_3)$. Their median \textit{associations} are all significant and monotonically increasing with cost quantile: $-0.31$, $-0.29$, $-0.09$, $0.35$ and $0.41$ for $\tau=0.10,0.25,0.50,0.75,0.90$, respectively. This suggest that, among larger banks (higher $\tau$) those who engage in off-balance sheet banking more heavily tend to enjoy scope economies of greater degrees. In contrast, for smaller banks (lower $\tau$), pivoting off balance sheet is associated with reduced scope-driven cost savings, plausibly because of their limited capabilities to capitalize on cross-output spillovers and input complementarities at smaller operations scales. 

% --------------------------------------------------
\begin{figure}[t]
	\centering
	\includegraphics[scale=0.6]{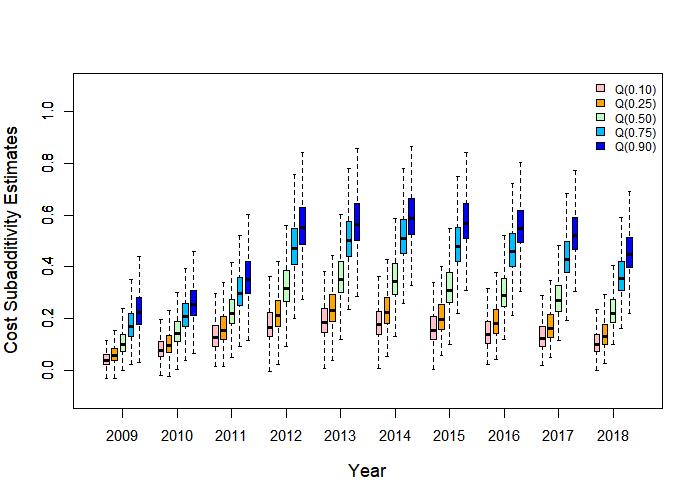}
	\caption{Evolution of Cost Subadditivity} \label{fig:boxplottime}
\end{figure}
% --------------------------------------------------

Lastly, we take a look at the evolution of scope economies. Figure \ref{fig:boxplottime} documents how distributions of the cost subadditivity estimates shifted over time. Plotted are the box-plots of $\mathcal{S}_t^*(\tau)$ across five considered cost quantiles $\tau$ for each year $t$. The data suggest a divergence in the degree of cost subadditivity between smaller (lower cost quantiles) and larger (higher cost quantiles) banks over time which, however, started reverting in the last years of the sample. We further observe that, while positive and significant throughout, the magnitude of a cost-saving potential associated with the product-scope diversification picked around 2013--2014 and has since been in a steady decline, across all cost quantiles.

% ------------------------------------------------------------------------------------------

\subsection{Scale Economies}

We complement our analysis of the scope-driven cost savings in the U.S.~banking with the examination of economies of scale. Scale economies are said to exist if the banks' average cost declines with equiproportional expansion of its outputs (i.e., with the increase in scale of production). As discussed in the introduction, the latter has been a subject of particular academic interest in face of the post-crisis regulatory reforms in the banking sector.

Our returns to scale measure takes into account quasi-fixity of the equity input per \citet{ caves1981productivity}:
\begin{equation}
\mathcal{R}_{t}(\tau)=\big(1-\partial\mathcal{Q}_c(\tau|\cdot)/\partial k_1\big)\Big/\sum_m\partial\mathcal{Q}_c(\tau|\cdot)/\partial y_m,
\end{equation}
where we replaced the usual $\log\mathcal{C}_{t}(\cdot)$ with the quantile function of the log-cost $\mathcal{Q}_c(\tau|\cdot)$ in the formula since our cost function estimation is for a conditional quantile. The measure of returns to scale is therefore both observation- and cost-quantile-specific.

Just like in the case of scope economies, for a given $\tau$, we are mainly interested in the following two hypotheses: (\textit{i}) $\mathbb{H}_0: \mathcal{R}_t(\tau)\le1\ \text{v.}\ \mathbb{H}_1: \mathcal{R}_t(\tau)>1$ and (\textit{ii}) $\mathbb{H}_0: \mathcal{R}_t(\tau)=1\ \text{v.}\ \mathbb{H}_1: \mathcal{R}_t(\tau)\ne1$. In case of (\textit{i}), rejection of the null would imply that the returns to scale statistically \textit{exceed} 1 implying increasing returns (IRS) and, thus, significant scale economies. In case of (\textit{ii}), failure to reject the null would suggest that returns to scale are statistically indistinguishable from 1, which is consistent with the bank exhibiting constant returns to scale (CRS) and, hence, scale invariance of costs.

% ---------------------------------------
\begin{table}[t]
\centering
\caption{ Returns to Scale Estimates}\label{tab:scale}
\footnotesize
\makebox[\linewidth]{
\begin{tabular}{lcccc|rr|rr}
\toprule
Cost & \multicolumn{4}{c}{\textit{Point Estimates}} 	& \multicolumn{4}{c}{\textit{Inference Categories, \%}} \\   
Quantiles ($\tau$)& Mean  & 1st Qu. & Median & 3rd Qu. & $\mathbf{=1}$ & $\ne1$  & $\mathbf{>1}$ & $\le1$ \\
			\midrule			
			$\mathcal{Q}(0.10)$  & 1.300 & 1.263 & 1.293 & 1.328 & \bf 0.02 & 99.98 & \bf 99.99 & 0.01 \\
			& (1.263, 1.352) & (1.226, 1.316) & (1.257, 1.344) & (1.286, 1.382) &       &      &       \\
			$\mathcal{Q}(0.25)$ & 1.321 & 1.282 & 1.313 & 1.351 & \bf 0.01 & 99.99 & \bf 100.0 & 0.00 \\
			& (1.282, 1.363) & (1.243, 1.322) & (1.276, 1.354) & (1.306, 1.399) &       &      &       \\
			$\mathcal{Q}(0.50)$  & 1.361 & 1.316 & 1.351 & 1.394 & \bf 0.01 & 99.99 & \bf 100.0 & 0.00 \\
			& (1.319, 1.404) & (1.276, 1.356) & (1.31, 1.393) & (1.347, 1.444) &       &      &       \\
			$\mathcal{Q}(0.75)$  & 1.405 & 1.352 & 1.392 & 1.441 & \bf 0.00 & 100.0 & \bf 100.0 & 0.00  \\
			& (1.352, 1.457) & (1.307, 1.398) & (1.344, 1.443) & (1.385, 1.500) &       &      &       \\
			$\mathcal{Q}(0.90)$  & 1.430 & 1.373 & 1.416 & 1.469 & \bf 0.01 & 99.99 & \bf 100.0 & 0.00  \\
			& (1.363, 1.491) & (1.314, 1.421) & (1.353, 1.469) & (1.397, 1.533) &       &      &       \\						
			\midrule
			\multicolumn{9}{p{14.6cm}}{\scriptsize The left panel summarizes point estimates of $\mathcal{R}_t(\tau)$ with the corresponding two-sided 95\% bias-corrected confidence intervals in parentheses. Each bank-year is classified as exhibiting IRS [$\mathcal{R}_t(\tau)>1$] vs.~non-IRS [$\mathcal{R}_t(\tau)\le1$] and CRS [$\mathcal{R}_t(\tau)=1$] vs.~non-CRS [$\mathcal{R}_t(\tau)\ne1$] using the corresponding one- and two-sided 95\% bias-corrected confidence bounds, respectively. The right panel reports sample shares for each category and for its corresponding negating alternative. Percentage points sum up to a hundred within binary groups only.} \\
			\bottomrule[1pt]
\end{tabular}
}
\end{table}
% ---------------------------------------

Table \ref{tab:scale} summarizes point estimates of the returns to scale for all estimated quantiles of the conditional cost distribution of banks. The right panel of the table reports the results of the hypothesis tests. Namely, reported is the breakdown of banks that exhibit IRS (scale economies) vs.~non-IRS (scale non-economies) and of banks that exhibit CRS (scale invariance) vs.~non-CRS (scale non-invariance).

The results in Table \ref{tab:scale} provide overwhelming evidence of ubiquitous scale economies in the banking sector, across all cost quantiles. The average point estimates of returns to scale ranges from 1.30 to 1.43, with banks from the higher quantiles of cost distribution exhibiting increasing returns to scale of larger magnitudes compared to those from the lower quantiles. We find that almost every single bank in our sample exhibits statistically significant scale economies (IRS). These results suggest that, when the bank radially expands the scale of its operation, its average variable cost decreases. These findings are consistent with the prior results which  however are almost exclusively based on the analyses of bank costs at the conditional \textit{mean} \citep[e.g.,][]{wheelock2012large,hughes2013said,restrepo2015nonparametric, malikov2015estimation,restrepo2015obelix, wheelock2018evolution}. Given that we find evidence of significant scale economies along the entire cost \textit{distribution}, our results provide the robust assurance to these earlier findings reported in the literature.

% ------------------------------------------------------------------------------------------

\subsection{Technological Change}

We conclude our analysis of bank cost structure by examining temporal shifts in the bank cost frontier in face of technological advancements as well as regulatory changes in the industry in aftermath of the 2008 financial crisis. A cost-diminishing technological change can provide another means for cost savings.

Because we model temporal variation in the cost relationship using discretized time indices, we replace the standard continuous measure of technical change with a discrete dual measure of technological change at each cost quantile $\tau$. Namely, from \eqref{eq:model_quant}, we have
\begin{align}
- \mathcal{TC}_{t}(\tau)\equiv&\ \mathcal{Q}_c(\tau|\cdot,t)-\mathcal{Q}_c(\tau|\cdot,t-1) \notag \\
=&\ \Delta L(t)+\Delta S(t)q_{\tau}\ + \notag \\ 
&\ \Big[\boldsymbol{\beta}_3\Delta L(t)+\boldsymbol{\gamma}_3 \Delta S(t)q_{\tau}\Big]'\boldsymbol{v}_{it} + \frac{1}{2}\Big[\boldsymbol{\beta}_4\Delta L(t)+\boldsymbol{\gamma}_4\Delta S(t)q_{\tau}\Big]'\text{vec}\left( \boldsymbol{v}_{it}\boldsymbol{v}_{it}'\right) ,
\end{align}
where $\Delta L(t)=L(t)-L(t-1)$ and $\Delta S(t)=S(t)-S(t-1)$, with its feasible analogue given by
\begin{align}\label{eq:tc}
-TC_{t}(\tau)=&\ \big(\eta_{\kappa}+\theta_{\kappa}q_{\tau}\big)D_{\kappa,t}-\big(\eta_{\kappa-1}+\theta_{\kappa-1}q_{\tau}\big)D_{\kappa-1,t-1}\ + \notag \\
&\ \Big[\big(\boldsymbol{\beta}_3\eta_{\kappa}+\boldsymbol{\gamma}_3\theta_{\kappa}q_{\tau}\big)D_{\kappa,t}-\big(\boldsymbol{\beta}_3\eta_{\kappa-1}+\boldsymbol{\gamma}_3\theta_{\kappa-1}q_{\tau}\big)D_{\kappa-1,t-1}\Big]'\boldsymbol{v}_{it} \ + \notag \\
&\ \frac{1}{2}\Big[\big(\boldsymbol{\beta}_4\eta_{\kappa}+\boldsymbol{\gamma}_4\theta_{\kappa}q_{\tau}\big)D_{\kappa,t} - \big(\boldsymbol{\beta}_4\eta_{\kappa-1}+\boldsymbol{\gamma}_4\theta_{\kappa-1}q_{\tau}\big)D_{\kappa-1,t-1}\Big]'\text{vec}\left[ \boldsymbol{v}_{it}\boldsymbol{v}_{it}'\right].
\end{align}
The first line in \eqref{eq:tc} corresponds to Hick-neutral component of technological change, whereas the last two lines represent non-neutral change.

The point estimates of technological change at different cost quantiles are summarized in Table \ref{tab:tc}. The right panel of the table reports results of a one-sided test of $\mathbb{H}_0: TC_t(\tau)\le0\ \text{v.}\ \mathbb{H}_1: TC_t(\tau)>0$, i.e., a test of whether $TC_t(\tau)$ is statistically positive implying that the bank enjoys technological \textit{progress} and, therefore, a \textit{ceteris paribus} cost diminution over time.

% ---------------------------------------
\begin{table}[t]
\centering
\caption{Technical Change Estimates}\label{tab:tc}
\footnotesize
\makebox[\linewidth]{
\begin{tabular}{lcccc|rrrr}
\toprule
Cost & \multicolumn{4}{c}{\textit{Point Estimates}} 	& \multicolumn{4}{c}{\textit{Inference Categories, \%}} \\  
Quantiles ($\tau$)& Mean  & 1st Qu. & Median & 3rd Qu. & $\mathbf{=0}$ & $\ne0$  & $\mathbf{>0}$ & $\le0$ \\
\midrule
$\mathcal{Q}(0.10)$ & --0.010 & --0.026 & --0.008 & 0.006 & \bf 63.06 & 36.94 & \bf 9.69 & 90.31\\
		& (--0.022, 0.004) & (--0.039, --0.010) & (--0.020, 0.006) & (--0.007, 0.021)  &   &   &  &      \\
$\mathcal{Q}(0.25)$ & --0.005 & --0.019 & --0.003 & 0.011  & \bf 61.23 & 38.77 & \bf 17.52 & 82.48 \\
		& (--0.015, 0.007) & (--0.031, --0.007) & (--0.013, 0.009) & (0.000, 0.025) &    &    &  &     \\
$\mathcal{Q}(0.50)$ & 0.005 & --0.008 & 0.007 & 0.021  & \bf 53.71 & 46.29 & \bf 36.91 & 63.09 \\
		& (--0.004, 0.015) & (--0.018, 0.001) & (--0.002, 0.018) & (0.011, 0.037) &       &      &  &   \\
$\mathcal{Q}(0.75)$ & 0.015 & 0.002 & 0.017 & 0.031 & \bf 47.22 & 52.78 & \bf 54.33 & 45.67 \\
		& (0.006, 0.026) & (--0.008, 0.013) & (0.008, 0.030) & (0.019, 0.049) &       &       &  &  \\
$\mathcal{Q}(0.90)$ & 0.021 & 0.007 & 0.022 & 0.037 & \bf 42.91 & 57.09 & \bf 61.50 & 38.50 \\
		& (0.008, 0.032) & (--0.005, 0.019) & (0.010, 0.035) & (0.022, 0.056) &       &     &  &   \\
	\midrule
\multicolumn{9}{p{15.5cm}}{\scriptsize The left panel summarizes point estimates of $TC_t(\tau)$ with the corresponding two-sided 95\% bias-corrected confidence intervals in parentheses. Each bank-year is classified as exhibiting technical progress [$TC_t(\tau)>0$] vs.~non-progress [$TC_t(\tau)\le0$] and technical stasis [$TC_t(\tau)=0$] vs.~non-stasis [$TC_t(\tau)\ne0$] using the corresponding one- and two-sided 95\% bias-corrected confidence bounds, respectively. The right panel reports sample shares for each category and for its corresponding negating alternative. Percentage points sum up to a hundred within binary groups only. } \\
\bottomrule[1pt]
\end{tabular}
	}
\end{table}
% ---------------------------------------

Our data suggest that, in the period following the financial crisis, only larger banks in the upper tail of the cost distribution ($\tau\ge0.75$) have been benefiting from significant cost-diminishing technological advances: 1.5--2.1\% p.a., on average. The share of these banks with statistically positive technological change estimates is between 54 and 62\%, with the rest of banks in the upper quartile exhibiting no statistically significant cost diminution. For mid-size banks at the median of the cost distribution, technical change is statistically positive for modest 37\% of banks. Evidence of significant cost diminution is even weaker among banks in the bottom half of the cost distribution.
Overall, our results suggest that the cost-saving effects of many recent technological advancements in the banking industry, such as the growing networks of automated teller machines, growing credit card networks, electronic payments, internet banking, etc., that were found in the pre-crisis period by earlier studies \citep[e.g.,][]{wheelock1999technical, almanidis2013accounting,malikov2015estimation} have now largely waned, plausibly because most banks had already capitalized on them to the fullest extent feasible and/or because they now face new regulatory controls. A significant technical change among larger banks in the upper tail of the cost distribution is likely due to their better capability to adapt and innovate.

% ------------------------------------------------------------------------------------------
% ------------------------------------------------------------------------------------------

\section{Conclusion}
\label{sec:conclusion}

Propelled by the recent financial product innovations, banks are becoming more complex, branching out into many ``nontraditional'' banking operations beyond issuance of loans. This broadening of operational scope in a pursuit of revenue diversification may be beneficial if banks exhibit scope economies. The existing empirical evidence lends no support for such product-scope-driven cost economies in banking, but it is greatly outdated and, surprisingly, there has been little (if any) research on this subject despite the drastic transformations that the U.S.~banking industry has undergone over the past two decades in the wake of technological advancements and regulatory changes. Commercial banks have significantly shifted towards nontraditional operations, and the portfolio of products offered by present-day banks is very different from that two decades ago. This underscore the importance of taking a fresh look at scope economies in banks because leveraging operational scope continues to play a vital role in operations management in banking. It is also important from a policy evaluation perspective, in the face of new financial regulations such as the Dodd–Frank Wall Street Reform and the Consumer Protection Act of 2010 that seek to set restrictions on the scale and scope of bank operations.

This paper provides new evidence about scope economies in U.S.~commercial banking during the 2009--2018 post-crisis period. We improve upon the prior literature not only by analyzing the most recent and relevant data and accounting for bank's nontraditional off-balance sheet operations, but also in multiple methodological ways as follows. In a pursuit of robust estimates of scope economies and statistical inference thereon, we estimate a flexible, yet parsimonious, time-varying-coefficient panel-data quantile regression model which accommodates three-way bank heterogeneity: (\textit{i}) distributional heterogeneity in the cost structure of banks along the size of their costs, (\textit{ii}) temporal variation in cost complementarities and spillovers due to technological change/innovation, and (\textit{iii}) unobserved bank confounders such as latent management quality. Our results provide strong evidence in support of significantly positive scope economies across banks of virtually all sizes. Contrary to earlier studies, we find no empirical corroboration for scope diseconomies.

% ------------------------------------------------------------------------------------------
% ------------------------------------------------------------------------------------------

{\small \setlength{\bibsep}{0pt} \bibliography{SPSCQcubib}}

% ------------------------------------------------------------------------------------------
% ------------------------------------------------------------------------------------------
\clearpage
\section*{Appendix}
\appendix

\section{Three-Step Estimation Procedure}
\label{sec:appx_estimator}

\setcounter{table}{0} 
\setcounter{figure}{0} 
\renewcommand\thetable{A.\arabic{table}} 
\renewcommand\thefigure{A.\arabic{figure}}

This appendix describes the estimation details of a conditional quantile function in \eqref{eq:model_quant_feasible}. First, for ease of notation, we define $\boldsymbol{D}_{t}=[D_{2,t},\dots,D_{T,t}]'$, $\boldsymbol{\eta}=[\eta_2,\dots,\eta_T]'$ and $\boldsymbol{\theta}=[\theta_2,\dots,\theta_T]'$.

\paragraph{Step 1.} We first estimate parameters of the location function. Under the assumption (\textit{ii}), from \eqref{eq:model_l} it follows that the conditional mean function of the log-cost $c_{it}$ is  
\begin{equation}\label{eq:model_step1*}
\mathbb{E}\left[c_{it}| \boldsymbol{v}_{it},\boldsymbol{D}_t\right]=\beta_{0}+\sum_{\kappa}\eta_{\kappa}D_{\kappa,t}+\Big[\boldsymbol{\beta}_{1}+\boldsymbol{\beta}_1^*\sum_{\kappa}\eta_{\kappa}D_{\kappa,t}\Big]'\boldsymbol{v}_{it}+ \frac{1}{2}\Big[\boldsymbol{\beta}_{2}+\boldsymbol{\beta}_2^*\sum_{\kappa}\eta_{\kappa}D_{\kappa,t}\Big]'\text{vec}\left( \boldsymbol{v}_{it}\boldsymbol{v}_{it}'\right) +\lambda_{i},
\end{equation}
which can be consistently estimated in the within-transformed form via nonlinear least squares after purging additive location fixed effects.\footnote{Note that, although \eqref{eq:model_step1*} is nonlinear, the presence of fixed effects does not give rise to the incidental parameter problem in this case because $\{\lambda_i\}$ enters the model additively and is not inside the nonlinear mean function.} 

Given the nonlinearity and high dimensionality of \eqref{eq:model_step1*} in parameters, we estimate the slope coefficients via concentration by noticing that, conditional on $\boldsymbol\eta$, this mean regression is linear in $\big[\boldsymbol{\beta}_{1}',\boldsymbol{\beta}_2',\boldsymbol{\beta}_1^*{'},\boldsymbol{\beta}_2^*{'}\big]'$ yielding the profiled least-squares estimator for $\big[\boldsymbol{\beta}_{1}(\boldsymbol\eta)',\boldsymbol{\beta}_2(\boldsymbol\eta)',\boldsymbol{\beta}_1^*(\boldsymbol\eta){'},\boldsymbol{\beta}_2^*(\boldsymbol\eta){'}\big]'$. Specifically, letting the concentrated sum of (within-transformed) squared errors be
\begin{align}\label{eq:step1_profile_rss}
M(\boldsymbol\eta)=\sum_{i}\sum_{t}\Bigg[& c_{it}-\overline{c}_{i}-\boldsymbol{\eta}'\big(\boldsymbol{D}_{t}-\overline{\boldsymbol{D}}\big)
 -\big(\boldsymbol{v}_{it}-\overline{\boldsymbol{v}}_{i}\big)'\boldsymbol{\beta}_{1}\left(\boldsymbol\eta\right)\ - \notag \\
&  \Big( \boldsymbol{\eta}'\boldsymbol{D}_t\cdot\boldsymbol{v}_{it}-\overline{\boldsymbol{\eta}'\boldsymbol{D}_t\cdot\boldsymbol{v}_{i}}\Big)'\boldsymbol{\beta}_{1}^*\left(\boldsymbol\eta\right)-  \frac{1}{2}\Big(\text{vec}\left( \boldsymbol{v}_{it}\boldsymbol{v}_{it}'\right)-\overline{\text{vec}\left( \boldsymbol{v}_{i}\boldsymbol{v}_{i}'\right)}\Big)\boldsymbol{\beta}_{2}\left(\boldsymbol\eta\right)\ - \notag \\
&  \frac{1}{2}\Big(\boldsymbol{\eta}'\boldsymbol{D}_t\cdot\text{vec}\left( \boldsymbol{v}_{it}\boldsymbol{v}_{it}'\right)-\overline{\boldsymbol{\eta}'\boldsymbol{D}_t\cdot\text{vec}\left( \boldsymbol{v}_{i}\boldsymbol{v}_{i}'\right)}\Big)'\boldsymbol{\beta}_{2}^*\left(\boldsymbol\eta\right)\Bigg]^{2},
\end{align}
with the ``bar'' denoting the cross-time averages of variables that it tops, we have the profiled estimators $[\boldsymbol{\beta}_{1}(\boldsymbol\eta)',\boldsymbol{\beta}_2(\boldsymbol\eta)', \boldsymbol{\beta}_1^*(\boldsymbol\eta){'},\boldsymbol{\beta}_2^*(\boldsymbol\eta){'}]' =\big(\sum_{i}\sum_{t}\mathbb{X}_{it}\mathbb{X}_{it}'\big)^{-1}\sum_{i}\sum_{t}\mathbb{X}_{it}\mathbb{Y}^{\dagger}_{it}$,  where $\mathbb{X}_{it}=\big[\left(\boldsymbol{v}_{it}-\overline{\boldsymbol{v}}_{i}\right)',$ $\big( \boldsymbol{\eta}'\boldsymbol{D}_t\cdot\boldsymbol{v}_{it}-\overline{\boldsymbol{\eta}'\boldsymbol{D}_t\cdot\boldsymbol{v}_{i}}\big)'$, $\frac{1}{2}\big(\text{vec}\left( \boldsymbol{v}_{it}\boldsymbol{v}_{it}'\right)-\overline{\text{vec}\left( \boldsymbol{v}_{i}\boldsymbol{v}_{i}'\right)}\big)$,  $\frac{1}{2}\big(\boldsymbol{\eta}'\boldsymbol{D}_t\cdot\text{vec}\left( \boldsymbol{v}_{it}\boldsymbol{v}_{it}'\right)-\overline{\boldsymbol{\eta}'\boldsymbol{D}_t\cdot\text{vec}\left( \boldsymbol{v}_{i}\boldsymbol{v}_{i}'\right)}\big)'$  and
$\mathbb{Y}_{it}^{\dagger} = c_{it}-\overline{c}_{i}-\boldsymbol{\eta}'\big(\boldsymbol{D}_{t}-\overline{\boldsymbol{D}}\big)$.
	
Thus, the nonlinear fixed-effects estimators of the slope coefficients $[\boldsymbol{\eta}',\boldsymbol{\beta}_{1}',\boldsymbol{\beta}_2',\boldsymbol{\beta}_1^*{'},\boldsymbol{\beta}_2^*{'}]'$ in the location functions are
\begin{equation}\label{eq:step1}
\widehat{\boldsymbol\eta}=\arg\min_{\boldsymbol\eta}M(\boldsymbol\eta)\quad \text{and}\quad \widehat{\boldsymbol{\beta}}_{1}=\boldsymbol{\beta}_{1}\left(\widehat{\boldsymbol\eta}\right),\ \quad \widehat{\boldsymbol{\beta}}_{2}=\boldsymbol{\beta}_{2}\left(\widehat{\boldsymbol\eta}\right),\
\quad \widehat{\boldsymbol{\beta}}_{1}^*=\boldsymbol{\beta}_{1}^*\left(\widehat{\boldsymbol\eta}\right),\
\quad \widehat{\boldsymbol{\beta}}_{2}^*=\boldsymbol{\beta}_{2}^*\left(\widehat{\boldsymbol\eta}\right).
\end{equation}
	
Under the usual $\sum_{i=1}^n\lambda_{i}=0$ normalization, we can then recover the location-shifting intercept ${\beta}_{0}$ and fixed effects $\{\lambda_i\}$ via
\begin{align}
\widehat{\beta}_{0}&=\frac{1}{nT}\sum_{i}\sum_{t}\Bigg(c_{it}-\widehat{\boldsymbol{\eta}}'\boldsymbol{D}_t -\Big[\widehat{\boldsymbol{\beta}}_{1}+\widehat{\boldsymbol{\eta}}'\boldsymbol{D}_t\cdot\widehat{\boldsymbol{\beta}}_1^*\Big]'\boldsymbol{v}_{it}-
\frac{1}{2}\Big[\widehat{\boldsymbol{\beta}}_{2}+\widehat{\boldsymbol{\eta}}'\boldsymbol{D}_t\cdot\widehat{\boldsymbol{\beta}}_2^*\Big]' \text{vec}\left( \boldsymbol{v}_{it}\boldsymbol{v}_{it}'\right) \Bigg), \label{eq:step1_1} \\
\widehat{\lambda}_{i}&=\frac{1}{T}\sum_{t}\Bigg(c_{it}-\widehat{\beta}_0-\widehat{\boldsymbol{\eta}}'\boldsymbol{D}_t -\Big[\widehat{\boldsymbol{\beta}}_{1}+\widehat{\boldsymbol{\eta}}'\boldsymbol{D}_t\cdot\widehat{\boldsymbol{\beta}}_1^*\Big]'\boldsymbol{v}_{it}-
\frac{1}{2}\Big[\widehat{\boldsymbol{\beta}}_{2}+\widehat{\boldsymbol{\eta}}'\boldsymbol{D}_t\cdot\widehat{\boldsymbol{\beta}}_2^*\Big]' \text{vec}\left( \boldsymbol{v}_{it}\boldsymbol{v}_{it}'\right) \Bigg)\ \forall i. \label{eq:step1_2}
\end{align}

Hence, the residual estimator is
\begin{equation}
\widehat{u}_{it}=c_{it}-\widehat{\beta}_0-\widehat{\boldsymbol{\eta}}'\boldsymbol{D}_t -\Big[\widehat{\boldsymbol{\beta}}_{1}+\widehat{\boldsymbol{\eta}}'\boldsymbol{D}_t\cdot\widehat{\boldsymbol{\beta}}_1^*\Big]'\boldsymbol{v}_{it}-
\frac{1}{2}\Big[\widehat{\boldsymbol{\beta}}_{2}+\widehat{\boldsymbol{\eta}}'\boldsymbol{D}_t\cdot\widehat{\boldsymbol{\beta}}_2^*\Big]' \text{vec}\left( \boldsymbol{v}_{it}\boldsymbol{v}_{it}'\right)-\widehat{\lambda}_{i}.
\end{equation}
	
\paragraph{Step 2.} We then estimate parameters of the scale function. Based on the assumptions (\textit{ii})--(\textit{iii}), we have an auxiliary conditional mean regression:
\begin{equation}\label{eq:model_2*}
\mathbb{E}\left[|u_{it}|| \boldsymbol{v}_{it},\boldsymbol{D}_t\right]=\gamma_{0}+\sum_{\kappa}\theta_{\kappa}D_{\kappa,t}+\Big[\boldsymbol{\gamma}_{1}+\boldsymbol{\gamma}_1^*\sum_{\kappa}\theta_{\kappa}D_{\kappa,t}\Big]'\boldsymbol{v}_{it}+ \frac{1}{2}\Big[\boldsymbol{\gamma}_{2}+\boldsymbol{\gamma}_2^*\sum_{\kappa}\theta_{\kappa}D_{\kappa,t}\Big]'\text{vec}\left( \boldsymbol{v}_{it}\boldsymbol{v}_{it}'\right) +\sigma_{i},
\end{equation}
which, just like in the first step, we can estimate via nonlinear least squares after within-transforming scale fixed effects out. 
Concretely, with the concentrated squared residual objective function 
\begin{align}\label{eq:step2_profile_rss}
M(\boldsymbol\theta)=\sum_{i}\sum_{t}\Bigg[& |\widehat{u}_{it}|-|\overline{\widehat{u}}_{i}|-\boldsymbol{\theta}'\big(\boldsymbol{D}_{t}-\overline{\boldsymbol{D}}\big)
-\big(\boldsymbol{v}_{it}-\overline{\boldsymbol{v}}_{i}\big)'\boldsymbol{\gamma}_{1}\left(\boldsymbol\theta\right)\ - \notag \\
&  \Big( \boldsymbol{\theta}'\boldsymbol{D}_t\cdot\boldsymbol{v}_{it}-\overline{\boldsymbol{\theta}'\boldsymbol{D}_t\cdot\boldsymbol{v}_{i}}\Big)'\boldsymbol{\gamma}_{1}^*\left(\boldsymbol\theta\right)-  \frac{1}{2}\Big(\text{vec}\left( \boldsymbol{v}_{it}\boldsymbol{v}_{it}'\right)-\overline{\text{vec}\left( \boldsymbol{v}_{i}\boldsymbol{v}_{i}'\right)}\Big)\boldsymbol{\gamma}_{2}\left(\boldsymbol\theta\right)\ - \notag \\
&  \frac{1}{2}\Big(\boldsymbol{\theta}'\boldsymbol{D}_t\cdot\text{vec}\left( \boldsymbol{v}_{it}\boldsymbol{v}_{it}'\right)-\overline{\boldsymbol{\theta}'\boldsymbol{D}_t\cdot\text{vec}\left( \boldsymbol{v}_{i}\boldsymbol{v}_{i}'\right)}\Big)'\boldsymbol{\gamma}_{2}^*\left(\boldsymbol\theta\right)\Bigg]^{2}, 
\end{align}
and the corresponding profiled estimators given by
$[\boldsymbol{\gamma}_{1}(\boldsymbol\theta)',\boldsymbol{\gamma}_2(\boldsymbol\theta)', \boldsymbol{\gamma}_1^*(\boldsymbol\theta){'},\boldsymbol{\gamma}_2^*(\boldsymbol\theta){'}]' =\big(\sum_{i}\sum_{t}\mathcal{X}_{it}\mathcal{X}_{it}'\big)^{-1}\times $ $\sum_{i}\sum_{t}\mathcal{X}_{it}\mathcal{Y}^{\dagger}_{it}$,  where $\mathcal{X}_{it}=\big[\left(\boldsymbol{v}_{it}-\overline{\boldsymbol{v}}_{i}\right)',$ $\big( \boldsymbol{\theta}'\boldsymbol{D}_t\cdot\boldsymbol{v}_{it}-\overline{\boldsymbol{\theta}'\boldsymbol{D}_t\cdot\boldsymbol{v}_{i}}\big)'$, $\frac{1}{2}\big(\text{vec}\left( \boldsymbol{v}_{it}\boldsymbol{v}_{it}'\right)-\overline{\text{vec}\left( \boldsymbol{v}_{i}\boldsymbol{v}_{i}'\right)}\big)$,  $\frac{1}{2}\big(\boldsymbol{\theta}'\boldsymbol{D}_t\cdot\text{vec}\left( \boldsymbol{v}_{it}\boldsymbol{v}_{it}'\right)-\overline{\boldsymbol{\theta}'\boldsymbol{D}_t\cdot\text{vec}\left( \boldsymbol{v}_{i}\boldsymbol{v}_{i}'\right)}\big)'$  and
$\mathcal{Y}_{it}^{\dagger} = |\widehat{u}_{it}|-|\overline{\widehat{u}}_{i}|-\boldsymbol{\theta}'\big(\boldsymbol{D}_{t}-\overline{\boldsymbol{D}}\big)$, the nonlinear fixed-effects estimators of the scale-function slope coefficients  $[\boldsymbol\theta',\boldsymbol{\gamma}_1',\boldsymbol{\gamma}_2',\boldsymbol{\gamma}_3',\boldsymbol{\gamma}_4']'$ are
\begin{equation}\label{eq:step2}
\widehat{\boldsymbol\theta}=\arg\min_{\boldsymbol\theta}M(\boldsymbol\theta)\quad \text{and}\quad \widehat{\boldsymbol{\gamma}}_{1}=\boldsymbol{\gamma}_{1}\left(\widehat{\boldsymbol\theta}\right),\ \quad \widehat{\boldsymbol{\gamma}}_{2}=\boldsymbol{\gamma}_{2}\left(\widehat{\boldsymbol\theta}\right),\
\quad \widehat{\boldsymbol{\gamma}}_{1}^*=\boldsymbol{\gamma}_{1}^*\left(\widehat{\boldsymbol\theta}\right),\
\quad \widehat{\boldsymbol{\gamma}}_{2}^*=\boldsymbol{\gamma}_{2}^*\left(\widehat{\boldsymbol\theta}\right).
\end{equation}
	
To recover the scale-shifting intercept ${\gamma}_{0}$ and fixed effects $\{\sigma_i\}$, use $\sum_{i=1}^n\sigma_i=0$:
\begin{align}
\widehat{\gamma}_{0}&=\frac{1}{nT}\sum_{i}\sum_{t}\Bigg(|\widehat{u}_{it}|-\widehat{\boldsymbol{\theta}}'\boldsymbol{D}_t -\Big[\widehat{\boldsymbol{\gamma}}_{1}+\widehat{\boldsymbol{\theta}}'\boldsymbol{D}_t\cdot\widehat{\boldsymbol{\gamma}}_1^*\Big]'\boldsymbol{v}_{it}-
\frac{1}{2}\Big[\widehat{\boldsymbol{\gamma}}_{2}+\widehat{\boldsymbol{\theta}}'\boldsymbol{D}_t\cdot\widehat{\boldsymbol{\gamma}}_2^*\Big]' \text{vec}\left( \boldsymbol{v}_{it}\boldsymbol{v}_{it}'\right) \Bigg), \label{eq:step2_1} \\
\widehat{\sigma}_{i}&=\frac{1}{T}\sum_{t}\Bigg(|\widehat{u}_{it}|-\widehat{\gamma}_0-\widehat{\boldsymbol{\theta}}'\boldsymbol{D}_t -\Big[\widehat{\boldsymbol{\gamma}}_{1}+\widehat{\boldsymbol{\theta}}'\boldsymbol{D}_t\cdot\widehat{\boldsymbol{\gamma}}_1^*\Big]'\boldsymbol{v}_{it}-
\frac{1}{2}\Big[\widehat{\boldsymbol{\gamma}}_{2}+\widehat{\boldsymbol{\theta}}'\boldsymbol{D}_t\cdot\widehat{\boldsymbol{\gamma}}_2^*\Big]' \text{vec}\left( \boldsymbol{v}_{it}\boldsymbol{v}_{it}'\right) \Bigg)\ \forall i. \label{eq:step2_2}
\end{align}

\paragraph{Step 3.}For any given quantile index $0<\tau<1$ of interest, we next estimate the unconditional quantile of $\varepsilon_{it}$. From \eqref{eq:model_s}, we have the conditional quantile function of $u_{it}$:
\begin{align}
\mathcal{Q}_{u}\left[\tau|\boldsymbol{v}_{it},\boldsymbol{D}_t\right]=
\Big(\gamma_{0}+\sum_{\kappa}\theta_{\kappa}D_{\kappa,t}+\Big[\boldsymbol{\gamma}_{1}+\boldsymbol{\gamma}_1^*\sum_{\kappa}\theta_{\kappa}D_{\kappa,t}\Big]'\boldsymbol{v}_{it}+ \frac{1}{2}\Big[\boldsymbol{\gamma}_{2}+\boldsymbol{\gamma}_2^*\sum_{\kappa}\theta_{\kappa}D_{\kappa,t}\Big]'\text{vec}\left( \boldsymbol{v}_{it}\boldsymbol{v}_{it}'\right)+\sigma_i\Big)q_{\tau}.
\end{align}

We therefore can estimate $q_{\tau}$ via a univariate quantile regression (with no intercept) via
\begin{equation} \label{eq:step3}
\widehat{q}_{\tau}=\arg\min_{q}\sum_{i}\sum_{t}\rho_{\tau}\Bigg\{\widehat{u}_{it}-\Big(\widehat{\gamma}_{0}+\widehat{\boldsymbol{\theta}}'\boldsymbol{D}_t+\Big[\widehat{\boldsymbol{\gamma}}_{1}+\widehat{\boldsymbol{\theta}}'\boldsymbol{D}_t\cdot\widehat{\boldsymbol{\gamma}}_1^*\Big]'\boldsymbol{v}_{it}+
\frac{1}{2}\Big[\widehat{\boldsymbol{\gamma}}_{2}+\widehat{\boldsymbol{\theta}}'\boldsymbol{D}_t\cdot\widehat{\boldsymbol{\gamma}}_2^*\Big]' \text{vec}\left( \boldsymbol{v}_{it}\boldsymbol{v}_{it}'\right)+\widehat{\sigma}_i\Big)q\Bigg\},
\end{equation}
where $\rho_{\tau}\{\xi\}=\xi\left(\tau-\mathbb{I}\left\{ \xi<0\right\} \right)$ is the check function, $\widehat{u}_{it}$ is estimated in Step 1, and $\big[\widehat{\boldsymbol\theta}',\widehat{\gamma}_{0}, \widehat{\boldsymbol{\gamma}}_{1}',\widehat{\boldsymbol{\gamma}}_{2}',\widehat{\boldsymbol{\gamma}}_{1}^*{'},\widehat{\boldsymbol{\gamma}}_{2}^*{'}\big]'$ and $\{\widehat{\sigma}_i\}$ are estimated in Step 2.
	
\bigskip
	
With all unknown parameters now estimated, we can construct the estimator of the feasible analogue of the $\tau$th conditional quantile of the log-cost in \eqref{eq:model_quant_feasible}.

% ------------------------------------------------------------------------------------------

\section{Bias-Corrected Bootstrap Inference}
\label{sec:inference}

\setcounter{table}{0} 
\setcounter{figure}{0} 
\renewcommand\thetable{B.\arabic{table}} 
\renewcommand\thefigure{B.\arabic{figure}}

To correct for finite-sample biases, we employ \citeauthor{efron1982}'s (1982) bias-corrected bootstrap percentile confidence intervals to conduct statistical inference. Bootstrap also significantly simplifies testing because, owing to a multi-step nature of our estimator, computation of the asymptotic variance of the parameter estimators is not trivial. Due to the panel structure of data, we use wild residual \textit{block} bootstrap, thereby taking into account the potential dependence in residuals within each bank over time. The bootstrap algorithm is as follows.
\begin{enumerate}\itemsep 0pt
\item[(i)] Compute the estimator in Step 1. Save the estimated coefficients  $\big[\widehat{\boldsymbol{\eta}}',\widehat{\beta}_0,\widehat{\boldsymbol{\beta}_{1}}',\widehat{\boldsymbol{\beta}}_2', \widehat{\boldsymbol{\beta}}_1^*{'},\widehat{\boldsymbol{\beta}}_2^*{'}\big]'$, location fixed effects $\{\widehat{\lambda}_i\}$ and
residuals $\{\widehat{u}_{it}\}$.

\item[(ii)] Generate bootstrap weights $\omega_{i}^b$ for each cross-section/bank $i$ from the two-point mass distribution:
\begin{equation}
w_{i}^{b}=\begin{cases}
\left(1+\sqrt{5}\right)/2 & \textrm{with prob.}\left(\sqrt{5}-1\right)/\left(2\sqrt{5}\right)\\
\left(1-\sqrt{5}\right)/2 & \textrm{with prob.}\left(\sqrt{5}+1\right)/\left(2\sqrt{5}\right)
\end{cases}.
\end{equation}
Next, for each observation $(i,t)$ with $i=1,\dots,n$ and $t=1,\dots,T$, generate a new bootstrap disturbance as $u_{it}^b = w_{i}^{b}\times\widehat{u}_{it}$.
	
\item[(iii)] Construct a new bootstrap outcome variable: 
$c_{it}^b=\widehat{\beta}_{0}+\sum_{\kappa}\widehat{\eta}_{\kappa}D_{\kappa,t}+\Big[\widehat{\boldsymbol{\beta}}_{1}+\widehat{\boldsymbol{\beta}}_1^*\sum_{\kappa}\widehat{\eta}_{\kappa}D_{\kappa,t}\Big]'\boldsymbol{v}_{it}+ \frac{1}{2}\Big[\widehat{\boldsymbol{\beta}}_{2}+\widehat{\boldsymbol{\beta}}_2^*\sum_{\kappa}\widehat{\eta}_{\kappa}D_{\kappa,t}\Big]'\text{vec}\left( \boldsymbol{v}_{it}\boldsymbol{v}_{it}'\right) +\widehat{\lambda}_{i}+u_{it}^{b}$ for all $i=1,\dots,n$ and $t=1,\dots,T$.

\item[(iv)] Recompute the Step 1 estimators in \eqref{eq:step1}--\eqref{eq:step1_2} using $ c_{it}^b$ in place of $ c_{it}$ to obtain bootstrap estimates of the location-function coefficients and fixed effects. Signify these by the superscript ``$b$.'' Then, compute the bootstrap estimate of the residual $\widehat{u}_{it}^{b}= c_{it}-\widehat{\beta}_{0}^{b}-\sum_{\kappa}\widehat{\eta}_{\kappa}^{b}D_{\kappa,t}-\Big[\widehat{\boldsymbol{\beta}}_{1}^{b}+\widehat{\boldsymbol{\beta}}_1^{*b}\sum_{\kappa}\widehat{\eta}_{\kappa}^{b}D_{\kappa,t}\Big]'\boldsymbol{v}_{it}-\allowbreak 
 \frac{1}{2}\Big[\widehat{\boldsymbol{\beta}}_{2}^{b}+\widehat{\boldsymbol{\beta}}_2^{*b}\sum_{\kappa}\widehat{\eta}_{\kappa}^{b}D_{\kappa,t}\Big]'\text{vec}\left[ \boldsymbol{v}_{it}\boldsymbol{v}_{it}'\right] -\widehat{\lambda}_{i}^{b}$. 

\item[(v)] Reestimate the Step 2 estimator in \eqref{eq:step2} and the Step 3 estimator in \eqref{eq:step3} using $\widehat{u}_{it}^{b}$ in place of $\widehat{u}_{it}$ to obtain bootstrap estimates of the scale function coefficients and $q_{\tau}$.
	
\item[(vi)] Repeat bootstrap steps (ii)--(v) $B$ times ($B=500$ in this study). Use the empirical distribution of $B$ bootstrap replicas of some estimand of interest (say, a coefficient or a quantile-specific function thereof such as cost subadditivity measure $\mathcal{S}_t^*$) to construct bias-corrected confidence intervals for this estimand.
\end{enumerate}

To make matters concrete, let the (potentially, observation- and quantile-specific) estimand of interest be denoted by $\widehat{\mathcal{E}}$. We can use the empirical distribution of $\{\widehat{\mathcal{E}}^1,\dots,\widehat{\mathcal{E}}^B\}$ to estimate the bias-corrected \textit{two}-sided $(1-\alpha)\times100$\% confidence bounds for $\mathcal{E}$ as an interval between the $[a_1\times100]$th and $[(1-a_2)\times100]$th percentiles of the bootstrap distribution, where $a_{1}=\Phi\left(2\widehat{z}_{0}+\Phi^{-1}(\alpha/2)\right) $ and $a_{2}=\Phi\left(2\widehat{z}_{0}+\Phi^{-1}(1-\alpha/2)\right)$ with $\Phi(\cdot)$ being the standard normal cdf along with its quantile function $\Phi^{-1}(\cdot)$. Parameter $ \widehat{z}_0 =  \Phi^{-1}\left(\#\big\{\widehat{\mathcal{E}}^b<\widehat{\mathcal{E}}\big\}/{B}\right)$ is a bias-correction factor measuring median bias, with $\#\{\mathcal{A}\}$ being a count function that returns the number of times event $\mathcal{A}$ is true. 
Naturally, to estimate the \textit{one}-sided lower/upper $(1-\alpha)\times100$\% confidence bound with bias correction, we respectively use the $[o_1\times100]$th or $[(1-o_2)\times100]$th percentiles of the bootstrap distribution, where $o_{1}=\Phi\left(2\widehat{z}_{0}+\Phi^{-1}(\alpha)\right) $ and $o_{2}=\Phi\left(2\widehat{z}_{0}+\Phi^{-1}(1-\alpha)\right)$.

% ------------------------------------------------------------------------------------------
% ------------------------------------------------------------------------------------------

\end{document}